\documentclass[12pt]{article}
\usepackage{graphicx}
\usepackage{amsmath}
\usepackage{amssymb} 
\usepackage{epstopdf}
\usepackage{url}
\usepackage[footnotesize]{caption}
\usepackage[usenames,dvipsnames]{color}
\usepackage[normalem]{ulem}

\begin{document}
\begin{titlepage}

\renewcommand{\thefootnote}{\fnsymbol{footnote}}

\begin{flushright}
 \today
\end{flushright}

\vspace{3em}

\begin{center}
 {\Large {\bf 
Tree-Level Unitarity and Renormalizability \\  \vspace{3mm} \ in Lifshitz Scalar Theory
 }}

 \vskip5em

 \setcounter{footnote}{0}
 {\sc Toshiaki Fujimori}$^1$,
 {\sc Takeo Inami}$^{2,3}$,
 {\sc Keisuke Izumi}$^4$, and
 {\sc Tomotaka Kitamura}$^5$


 \vskip3em

 
 $^1${\it 
 Department of Physics, and Research and Education Center for Natural
 Sciences, \\ Keio University, Hiyoshi 4-4-1, Kanagawa 223-8521, Japan
 } \vspace{1mm}

 $^2${\it 
 Department of Physics, National Taiwan University, Taipei, Taiwan
 } \vspace{1mm}

 $^3${\it 
Riken Nishina Center, Saitama, Japan
 } \vspace{1mm}

 $^4${\it 
Departament de F{\'\i}sica Fonamental, Institut de Ci\`encies del Cosmos (ICCUB),
\\
Universitat de Barcelona, Mart\'{\i} i Franqu\`es 1, E-08028 Barcelona, Spain
 } \vspace{1mm}
 
 $^5${\it 
Department of physics, Waseda university, Shinjuku, Tokyo, Japan
 }
 
 \vskip3em

\end{center}

 \vskip2em

\renewcommand{\baselinestretch}{1.4}\selectfont
\begin{abstract}
We study unitarity and renormalizability in the Lifshitz scalar field theory, 
which is characterized by an anisotropic scaling between the space and time directions. 
Without the Lorentz symmetry, 
both the unitarity and the renormalizability conditions are modified 
from those in relativistic theories. 
We show that for renormalizability, 
an extended version of the power counting condition is required 
in addition to the conventional one. 
The unitarity bound for S-matrix elements also gives stronger constraints on interaction terms  
because of the reference frame dependence of scattering amplitudes.  
We prove that both unitarity and renormalizability require identical conditions 
as in the case of conventional relativistic theories.  
\end{abstract}

\end{titlepage}
\tableofcontents
\newpage
\renewcommand{\thesubsubsection}{\roman{subsubsection}.}
\renewcommand{\baselinestretch}{1.4}\selectfont

\section{Introduction}
A field theory has to posses both renormalizability and unitarity 
to be a well-defined quantum theory. 
These properties, which may be intimately related to each other, 
are believed to provide identical constraints on consistent interactions.
This idea motivated the study of the equivalence 
between the renormalizability and the tree-level unitarity in gauge theories 
\cite{Llewellyn Smith:1973ey,Cornwall:1973tb,Cornwall:1974km}. 
It has been shown that a similar equivalence also holds for gravity theories \cite{Berends:1974gk}
and no counter-example is known in relativistic field theories\footnote{ 
Theories with negative norm states (ghost states), such as in higher curvature theory, 
are excluded in the discussion of the equivalence between unitarity and renormalizability.}.
It is worth checking whether the equivalence holds true for more generic field theories, such as non-relativistic theories. 
The purpose of this paper is to investigate the equivalence between renormalizability and unitarity 
in Lifshitz-type theories \cite{Lifshitz scalar}, which are characterized by the Lifshitz scaling 
 \begin{eqnarray}
t \to b^z t,  \hspace{2cm} x^i \to bx^i \hspace{5mm} (i=1,\dots d). \label{1}
\end{eqnarray}
Due to this anisotropic scaling property, 
Lifshitz-type theories have improved ultraviolet (UV) behaviors but lack the Lorentz symmetry.

Lifshitz-type theories have gotten a lot of attention in cosmology \cite{Izumi:2010yn} and 
some their quantum properties are also investigated \cite{Fujimori:2015wda,Anselmi:2007ri,Visser:2009fg,Colombo:2014lta}. 
Moreover, by introducing the idea of the Lifshitz scaling to gravity theory, 
Ho\v{r}ava has constructed a power-counting renormalizable gravity theory, 
which is called the Ho\v{r}ava-Lifshitz (HL) gravity \cite{Horava:2009uw}. 
The application of the HL gravity is widely studied in cosmology (see \cite{Mukohyama:2010xz} for a review), 
such as the emergence of the dark matter as an integration constant \cite{DM} and so on.
In contrast, the quantum properties, such as unitarity and renormalizability, of HL gravity still remain uncertain. 

In this paper, we study the quantum aspects of the Lifshitz scalar field theory, 
which can be viewed as a toy model of the HL gravity. 
The lack of the Lorentz symmetry gives rise to significant modifications to both renormalizability and unitarity. 
We show that due to the modification of the scaling dimensions of fields, 
an extended version of the power counting argument is necessary for renormalizability. 
As for tree-level unitarity, 
the reference frame dependence of scattering amplitudes 
gives rise to new modified constraints, which can be stronger than 
those derived only from the high energy center-of-mass (COM) scattering amplitudes. 
We derive the extended power-counting-renomalizability (PRC) conditions and the tree-level unitarity constraints 
on self-interactions of a single Lifshitz scalar field. 
Our result shows that the equivalence holds true 
also for the extended versions of renormalizability and unitarity. 

This paper is organized as follows. 
In Sec.\,\ref{Lfr}, we derive the extended PCR constraints, 
which are necessary conditions for renormalizability. 
After reviewing the unitarity bound in Sec.\,\ref{Secub}, 
we derive unitarity conditions for high-energy two-body scattering amplitudes 
of the Lifshitz scalar in Sec.\,\ref{UbLs}.
In Sec.\,\ref{Sec:const}, unitarity constraints on quartic and cubic interactions are derived 
from the unitarity conditions for the high-energy scattering amplitude. 
We see that the unitarity constraints are equivalent to the extended PCR constraints derived in Sec.\,\ref{Lfr}.
In Sec.\,\ref{SecTUOS}, s-channel scattering amplitudes with on-shell intermediate particles are discussed.
By correctly taking into account the resonance effects, 
we show that interactions satisfying the extended PCR conditions do not violate the unitarity bound. 
Finally, we give the summary and discussion in Sec.\,\ref{summary}.

\section{Renormalizability Condition}\label{Lfr}

In this section, we derive necessary conditions for the renormalizability of the Lifshitz scalar field theory. 
Since scalar fields can have negative scaling dimensions,
we have to carefully use the power counting argument to check the renormalizability of the theory. 
We show that renormalizable interactions 
have to satisfy an extended version of the PCR condition. 

The generic form of the action for a Lifshitz scalar field $\phi$ is given by 
\begin{eqnarray}
S = \int dt d^d x \bigg[ \frac{1}{2}\phi \Big\{ \partial_t^2 - f(-\triangle) \Big\} \phi + {\cal{L}}_{int} \bigg], 
\qquad \Delta:= \partial^i\partial_i, 
\label{2nd}
\end{eqnarray}
where $f(-\triangle)= (-\triangle)^z + \cdots$ is a polynomial of degree $z$ and 
the interaction Lagrangian ${\cal{L}}_{int}$ will be specified shortly.  
The dispersion relation takes the form
\begin{eqnarray}
E = \sqrt{f(p^2)},
\label{DR}
\end{eqnarray}
where $E$ and $p$ are the energy and the magnitude of the momentum, respectively.
This dispersion relation becomes $E \approx p^z$ in the UV regime
and is consistent with the Lifshitz scaling (\ref{1}). 
Since the asymptotic behavior of propagators is given by 
\begin{eqnarray}
\frac{1}{E^2 - f(p^2)} \approx \frac{1}{E^2-p^{2z}},
\end{eqnarray}
UV divergences of loop diagrams are milder 
than those in relativistic theories. 
The scaling dimension of $\phi$ can be read from the quadratic term in (\ref{2nd}):
\begin{eqnarray}
[\phi] = \frac{d-z}{2},
\label{eqphi}
\end{eqnarray}
where we have used the convention such that $[E]=z$ and $[p]=1$. 
In the following, we will consider the scaling law (\ref{1}) 
for arbitrary values of $z$ including $z>d$, for which $[\phi]$ is negative. 
We will see that the power counting argument should be modified for $[\phi] < 0$ ($z>d$). 
In our previous paper~\cite{Fujimori:2015wda}, 
we have investigated that the Lifshitz scalar theory with $z=d$, for which $[\phi] =0$, 
already presents unconventional features concerning its renormalizability.  

\subsection{Extended Power-Counting-Renormalizability Condition }
Let us consider the following generic $n$-point interaction term
\begin{eqnarray}
S_{int} = \lambda \int dt d^d x \, (\partial_x^{a_1} \phi) \, (\partial_x^{a_2} \phi) \cdots (\partial_x^{a_n} \phi), 
\label{eq:vertex}
\end{eqnarray}
where $a_1,\cdots, a_n$ are non-negative integers. 
Each $\partial_x$ denotes any of the spatial derivatives $\partial_i~(i=1,\cdots,d)$ 
and we suppose that this interaction term is designed 
to keep the $d$-dimensional spatially rotational symmetry $O(d)$.  
Without loss of generality, we can rearrange $\partial_x^{a_i} \phi~(i=1,\cdots,n)$ 
so that $0 \leq a_1 \leq a_2 \leq \cdots \leq a_n$. 
The dimension of the coupling constant $\lambda$ is given by
\begin{eqnarray}
[\lambda] = z + d - \sum_{l=1}^n ( a_l + [\phi] ). 
\label{eqlam}
\end{eqnarray}
In relativistic theories, 
an interaction is renormalizable if it has a coupling constant with non-negative dimension: $[\lambda] \geq 0$.
On the other hand, in the Lifshitz scalar theory, 
$[\lambda]$ has to satisfy a more strict condition, which can be derived as follows. 

For a one-particle irreducible (1PI) loop diagram, 
let $V_e~(e=0,1,\cdots,n-2)$ be the numbers of vertices with $e$ external lines. 
\begin{figure}[tb]
  \begin{center}
    \includegraphics[width=100mm]{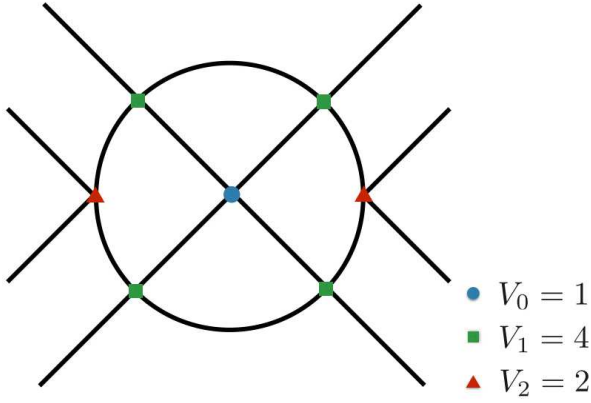}
  \end{center}
  \caption{A loop diagram with $n=4$, $V_0=1$, $V_1=4$, $V_2=2$ ($N=8$).}
  \label{fig:1PI}
\end{figure}
Note that $V_{n-1}=V_n=0$ for any 1PI loop diagrams. 
An example with $n=4$ is shown in Fig.\,\ref{fig:1PI}. 
The contribution of the vertex with $e$ external lines takes the form
\begin{eqnarray}
\mbox{Vertex for Eq.\,\eqref{eq:vertex}} ~=~ \lambda \, ( q_1^{a_1} \cdots q_e^{a_e} ) \, (p_1^{a_{e+1}} \cdots p_{n-e}^{a_n}) + (\mbox{permutations w.r.t. }a_i), 
\end{eqnarray}
where $p_i$ and $q_i$ are internal and external momenta, respectively. 
Since we have set $0 \leq a_1 \leq a_2 \leq \cdots \leq a_n$, 
the first term is the dominant contribution in the limit $p_i \rightarrow \infty$.
From this expression, 
the leading order part of the loop integral can be estimated as
\begin{eqnarray}
\mbox{1PI loop diagram} ~\approx~\int ( d E d^d p )^L \ \left( \frac{1}{E^2 - p^{2z}} \right)^P \ \prod_{e=0}^{n-2} \left( \prod_{l=e+1}^{n} p^{a_l} \right)^{V_e}, 
\end{eqnarray}
where $L$ and $P$ are the numbers of loops and internal propagators, respectively.
This implies that the degree of divergence of the loop diagram is 
\begin{eqnarray}
D = L (z+d) - 2 z P + \sum_{e=0}^{n-2} V_e \sum_{l=e+1}^{n} a_l. 
\label{eq:degree}
\end{eqnarray}
Since each vertex has $n-e$ internal lines and 
each internal line is connected to two vertices,  
the number of the internal propagators $P$ is given by
\begin{eqnarray}
P = \frac{1}{2} \sum_{e=0}^{n-2} V_e (n - e).
\end{eqnarray}
The number of loop integrals $L$ can be identified with the number of internal momenta
which are not determined by the momentum conservation law. 
Since there is one delta function at each vertex
and one of them is for the overall momentum, 
the number of loop $L$ is given by
\begin{eqnarray}
L = P - \sum_{e=0}^{n-2} V_e + 1.
\end{eqnarray}
Eliminating $L$ and $P$ from Eq.\,\eqref{eq:degree} and 
using Eqs.\,(\ref{eqphi}) and (\ref{eqlam}), 
we obtain 
\begin{eqnarray}
D ~=~ z+d - \sum_{e=0}^{n-2} V_e \big( \, [\lambda] \, + \, d_e \big), \hspace{10mm} 
d_e := \sum_{l=1}^e ( a_l + [\phi] ). 
\end{eqnarray}

\begin{figure}[tb]
\begin{center}
\includegraphics[width=100mm]{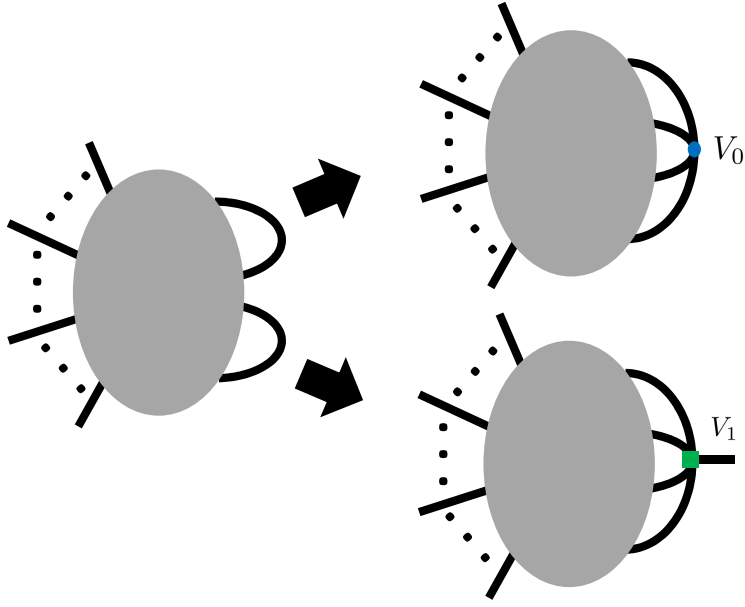}
\end{center}
\caption{Adding a vertex to divergent diagram: The left diagram is a divergent loop diagram. 
The upper and lower arrows show the operations of adding $V_0$ and $V_1$ vertices, respectively. 
The upper operation does not change the structures of external line, while the lower operation creates an additional external line.}
\label{fig:PCR}
\end{figure}

For a renormalizable interaction,  
$N$-point diagrams have to be finite for sufficiently large $N$ 
so that all UV divergences can be eliminated by a finite number of counter terms. 
In other words, $D$ has to be negative for loop diagrams with sufficiently large numbers of external lines $N = \sum_e e V_e$. 
If the contribution of each vertex $[\lambda]+d_e$ is negative for some $e$, 
the degree $D$ does not become negative as $V_e \rightarrow \infty$. 
It means that there are infinitely many UV divergent graphs with different $N$. 
Such divergences cannot be absorbed by a finite number of counter terms, 
and hence this type of interactions is non-renormalizable. 

\begin{figure}[tb]
  \begin{center}
    \includegraphics[width=100mm]{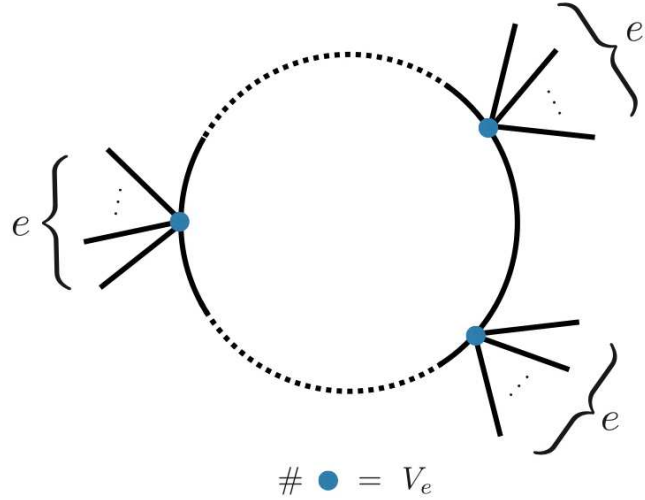}
  \end{center}
  \caption{A loop diagram constructed only with vertices satisfying $[\lambda]+d_e=0$:
Regardless of the number of vertices, this loop diagram diverges with degree $D = z+d$.}
  \label{fig:1loopVe}
\end{figure}

Next, we consider interactions with $[\lambda]+d_e = 0$ for some $e \ge 1$. 
The corresponding vertices in a loop diagram do not change the value of $D$ 
but increase the number of external lines (see Fig.\,\ref{fig:PCR}). 
Using this type of interactions, 
we can construct an infinite number of divergent loop diagrams 
with different numbers of external legs $N$ as shown in the example in Fig.\,\ref{fig:1loopVe}.  
Hence this type of interactions is also non-renormalizable. 
For $e=0$, in contrast, interactions with $[\lambda]+d_0 = 0 (=[\lambda]) $ are not excluded, 
since the addition of the corresponding vertices does not change the number of the external lines (see Fig.\,\ref{fig:PCR}).  
Therefore, as in the case of the conventional scalar field theory, 
the number of the required counter terms does not increase
and hence interaction terms with $[\lambda]=0$ can be renormalizable. 

In summary, the power-counting argument gives the following conditions 
for the coupling constant $\lambda$ of a renormalizable interaction: 
\begin{eqnarray}
&&[\lambda] ~ \ge \ \, \, 0, \hspace{12mm} (\mbox{for $e=0$}), \label{pcr} \\
&&[\lambda] ~ > -d_e, \hspace{10mm} (\mbox{for all $e=1,2,\cdots,n-2$}). 
\label{eq:renormalization_condition}
\end{eqnarray}
The first condition is the conventional PCR condition. 
In addition, the other conditions\,\eqref{eq:renormalization_condition} 
are also required for the renormalization in the Lifshitz scalar field theory. 
We call these conditions the extended PCR.
For $z \le d$, the dimension of the scalar field $[\phi]=(d-z)/2$ is non-negative
and hence $d_e~(e=1,\cdots,n-2)$ are also non-negative. 
In such a case, the conditions\,\eqref{eq:renormalization_condition}  
are automatically satisfied as long as the conventional PCR condition $[\lambda] \geq 0$ holds. 
On the other hand,  
$d_e~(e=1,\cdots,n-2)$ can be negative for $z > d$, 
so that $[\lambda]$ should satisfy the most strict of the conditions in \eqref{eq:renormalization_condition}. 

It is convenient to rewrite Eqs.\,\eqref{pcr} and \eqref{eq:renormalization_condition} into the following equivalent conditions: 
\begin{eqnarray}
\sum_{~l=1~}^n (a_l + [\phi]) &\le& z + d, \label{PCR}\\
\sum_{l=e+1}^n (a_l + [\phi]) &<&  z + d, \hspace{10mm} (\mbox{for all $e=1,\cdots,n-2$}). 
\label{eq:ePCR}
\end{eqnarray}
The latter condition implies that the dimension of 
$\partial_x^{a_{e+1}} \phi \cdots \partial_x^{a_n} \phi~(e=1,\cdots,n-2)$
should be less than $z+d$. 

To show the consistency of the extended PCR condition, 
it is important to check the renormalizability of counter terms introduced to cancel the UV divergences. 
In Appendix \ref{appendix:counter}, 
we show that the counter terms for renormalizable interactions also satisfy the extended PCR condition. 
If the dimension of the field is negative, 
there are an infinite number of terms satisfying the extended PCR. 
However, we show in Appendix \ref{Apfinite} that 
if the number of terms initially introduced in the bare action is finite, 
the theory can be renormalized with a finite number of counter terms.

\subsection{Example }
We illustrate the importance of 
the extended PCR condition \eqref{eq:renormalization_condition}  by taking an example.  
Due to the Lifshitz scaling, the scalar field has a negative dimension for $z>d$. 
In such cases, there can be interactions that have a coupling constant $\lambda$ 
with non-negative dimension but do not satisfy the condition \eqref{eq:renormalization_condition}. 
By using the corresponding vertices, we can construct infinitely many divergent diagrams
even if the conventional PCR condition \eqref{pcr} is satisfied. 

We consider an example with $d=3$ and $z=5$, 
in which the dimension of the scalar field is $[\phi] = -1$. 
The quartic interaction term (corresponding to $a_1=a_2=0,\  a_3=a_4 = 6$)
\begin{eqnarray}
S_4 = \lambda \int dt d^3x \, \phi^2 ( \Delta^3 \phi )^2, 
\label{4th}
\end{eqnarray}
has the dimensionless coupling constant $\lambda$, 
and thus this interaction satisfies the conventional PCR condition $[\lambda] \geq 0$. 
However, since $a_3 + a_4 + 2 [\phi] = 10 > 8 = z + d$, 
the extended PCR \eqref{eq:ePCR} with $e=2$ is not satisfied. 

\begin{figure}[tb]
  \begin{center}
    \includegraphics[keepaspectratio=true,height=60mm]{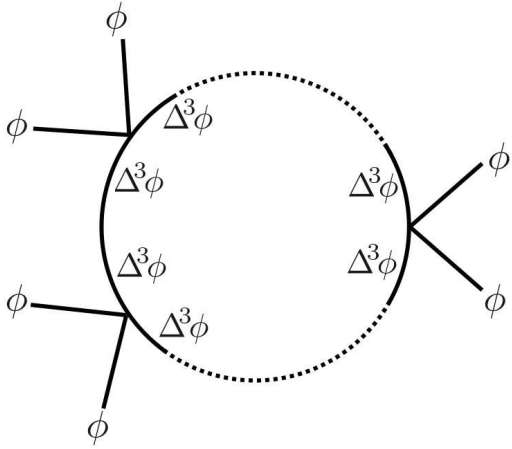}
  \end{center}
  \caption{$2n$-point one-loop diagram with $n$ 4-point vertices: 
  The dotted parts contain the same vertices as those drawn in the figure. 
  $\phi$ and $\Delta^3 \phi$ indicate that the neighboring lines correspond to $\phi$ and $\Delta^3 \phi$ in (\ref{4th}).}
  \label{1loop}
\end{figure}

Let us focus on the one-loop diagram with $2n$ external legs shown in Fig.\,\ref{1loop}. 
Its leading behavior is given by the term for which two ($\Delta^3 \phi$)'s 
at each vertex are contracted with those in the two neighboring vertices.
Thus, each vertex gives the leading behavior $\approx p^{12}$ 
for large loop momentum $p \rightarrow \infty$.
Since this diagram consists of $n$ four-point vertices and $n$ internal propagators, 
the degree of divergence is estimated as
\begin{eqnarray}
\int d\omega d^3 p \left(\frac{1}{\omega^2-p^{10}}\right)^n \left( p^{12}\right)^n
\sim \Lambda^{8+2n},
\end{eqnarray}
where $\Lambda$ is a UV cutoff with $[\Lambda]=1$. 
For any $n$, this loop integral diverges as $\Lambda \to \infty$, 
and thus the interaction \eqref{4th} is not renormalizable. 

The reason why the quartic action \eqref{4th} is non-renormalizable can be explained as follows. 
In the leading contribution of each vertex, 
the part of the operator \eqref{4th} involved in the loop integral is $(\Delta^3 \phi)^2$, 
whose dimension is $[(\Delta^3 \phi)^2] = 10$. 
Since the dimension of $\phi$ is negative, 
this is the higher than the dimension of the whole operator: $[\phi^2 (\Delta^3 \phi)^2]=8$. 
In general, operators with dimension $8(=z+d)$ represent marginal interactions 
while those with higher dimensions are non-renormalizable. 
Since the operator $(\Delta^3 \phi)^2$, 
which is relevant to the loop integral, 
has higher dimensions, 
the interaction \eqref{4th} is non-renormalizable. 
This peculiar phenomenon occurs because, if the dimension of the field $[\phi]$ is negative, 
a part of interaction term $(\Delta^3 \phi)^2$ can have higher dimensions 
than whole interaction term $\phi^2(\Delta^3 \phi)^2$; 
the same phenomenon does not occur in the conventional relativistic theory, 
in which $\phi$ always has a non-negative dimension. 

\section{Unitarity Bound}\label{Secub}
In this section, we briefly review the unitarity bound.  
It is a necessary condition for scattering amplitudes derived from the unitarity of the S-matrix
\begin{eqnarray}
SS^\dagger = \mathbf 1 .
\label{defM}
\end{eqnarray}
Decomposing the S-matrix as $S = \mathbf 1 + i T$, 
we can rewrite the unitarity condition into 
\begin{eqnarray}
-i(T - T^{\dagger})=TT^{\dagger}.
\label{T1}
\end{eqnarray}
We define the scattering amplitude ${\cal{M}}{\left(i \to f\right)}$ 
from the matrix element of $T$
by factoring out the delta functions
which indicate the energy-momentum conservation: 
\begin{eqnarray}
\langle f |T| i \rangle = \delta(E_i-E_f) \delta^{d} \left( {\bf p}_i - {\bf p}_f \right) {\cal{M}} \left( i \to f \right).
\label{defM}
\end{eqnarray}
Let $\{ |{X} \rangle \}$ be a complete orthonormal basis of the Hilbert space:
\begin{eqnarray}
\sum_{X}|X\rangle \langle X|=1.
\end{eqnarray} 
Then, Eq.$(\ref{T1})$ can be rewritten as 
\begin{eqnarray}
-i[{\cal{M}}\left(i\to f\right)-{\cal{M}}\left(f\to i \right)^{\ast}] = \sum_{X}\delta(E-E_X) \delta^{d} \left( {\bf p} - {\bf p}_X \right){\cal{M}}\left(i\to X\right){\cal{M}}\left(f\to X\right)^{\ast},
\end{eqnarray}
where $E(=E_i=E_f)$ and ${\bf p}(={\bf p}_{i}={\bf p}_{f})$ are the total energy and momentum respectively. 
In particular, for identical initial and final states, the relation reduces to the optical theorem
\begin{eqnarray}
2{\rm{Im}}{\cal{M}}\left(i\to i\right)=\sum_{X}\delta(E-E_X) \delta^{d} \left( {\bf p} - {\bf p}_X \right)|{\cal{M}}\left(i\to X\right)|^2,
\label{op}
\end{eqnarray}
where $\sum_{X}$ denotes the sum with respect to all possible intermediate states. 
In the following, we ignore unimportant numerical factors.

The $n$-particle Hilbert space is spanned by the standard 
asymptotic momentum eigenstates $|\bf{p}_1, \cdots,\bf{p}_n\rangle$, 
which we normalize as
\begin{eqnarray}
\int {\prod^n_{j=1}}\frac{d^{d} p_j}{2E_j}|{\bf{p}}_1, \cdots,{\bf{p}}_n \rangle \langle{\bf{p}}_1, \cdots,{\bf{p}}_n|=1,
\end{eqnarray}
with $E_j=\sqrt{f(p_j^2)}$.
In the discussion of the unitarity bound, 
it is convenient to use another basis $|E, {\bf{p}}, l\rangle$ with a discrete label $l$ defined as follows. 
First, let us consider the subspace of the $n$-particle momentum space 
specified by fixed values of the total energy and momentum 
\begin{eqnarray}
{\sum^n_{j=1}}E_j=E, \hspace{5mm} {\sum^n_{j=1}}{\bf{p}}_j = \bf{P}.
\end{eqnarray}
This constant-energy-momentum space, which we denote by $\mathcal C_{(E,\mathbf P)}$, 
is a compact subspace of $\mathbb R^{n(d+1)}$. 
Introducing a set of orthonormal functions $\{h_l\left({\bf p}_j \right)\}$ on $\mathcal C_{(E,\mathbf P)}$, 
we can define $|E,{\bf P},l\rangle$ by
\begin{eqnarray}
|E,{\bf P},l\rangle = \int{d\Pi_n} \, h_l\left( {\bf p}_j \right)|{{\bf p}}_1, \cdots{\bf{p}}_n\rangle, 
\label{epl}
\end{eqnarray}
where $d\Pi_n$ is the standard phase space volume element on $\mathcal C_{(E,\mathbf P)}$:
\begin{eqnarray}
{d\Pi_n}\equiv{\prod^n_{j=1}}\frac{d^dp_{j}}{2E_{j}}\delta \left({E_1}+\cdots+{E_n}-E\right)\delta^d\left({\bf{p}}_1+\cdots+{{\bf{p}}_n}-{\bf{P}}\right).\label{pi}
\end{eqnarray}
Let ${\cal{M}}\left(E,{\bf{P}};{l} \to {l'}\right)$ be the scattering amplitude defined by
\begin{eqnarray}
\langle E,{\bf{P}},{l}|T|E',{\bf{P'}},{l'}\rangle = \delta(E-E')\delta^d ({\bf {P}}-{\bf{P'}}){\cal{M}}\left(E,{\bf{P}};{l} \to {l'}\right). 
\end{eqnarray}
For this scattering amplitude, the unitarity bound can be obtained from Eq.(\ref{op}): 
\begin{eqnarray}
|{\cal{M}}\left(E,{\bf{P}};{l} \to {l}\right)| &\geq&|{\rm{Im}}{\cal{M}}\left(E,{\bf{P}};{l} \to {l}\right)| \phantom{\bigg(} \nonumber \\
&=&\sum_{l'}|{\cal{M}}\left(E,{\bf{P}};{l} \to {l'}\right)|^2 \geq |{\cal{M}}\left(E,{\bf{P}};{l} \to {l'}\right)|^2.
\label{eq:unitary_bound}
\end{eqnarray}
Here, in the last expression, we have picked up a specific state $l'$ 
from the sum over any possible intermediate states. 
If we choose $l'=l$, the unitary bound \eqref{eq:unitary_bound} implies that 
$|{\cal{M}}\left(E,{\bf{P}};{l} \to {l}\right)| < \text{const}$.
From this inequality and Eq.\,\eqref{eq:unitary_bound}, it follows that
\begin{eqnarray}
|{\cal{M}}\left(E,{\bf{P}};{l} \to {l'}\right)| \leq \text{const}, \hspace{10mm} \forall \ l, l'.
\end{eqnarray}
Therefore, unitarity is violated 
if the scattering amplitudes between the normalized discrete states \eqref{epl} 
are not bounded from above.

For later use, we rewrite ${\cal{M}}\left(E,{\bf{P}};{l} \to {l'}\right)$ 
in terms of the scattering amplitudes between the standard momentum eigenstates: 
\begin{eqnarray}
&& \langle {\bf{p}}_1,\cdots,{\bf{p}}_{n_i} |T| {\bf{k}}_1,\cdots,{\bf{k}}_{n_f}\rangle \phantom{\Big[} \nonumber \\
&&\qquad
= \delta \Bigg(\sum_{j=1}^{n_i} E_j - \sum_{m=1}^{n_f} E_m \Bigg) \delta^{d}\left(\sum_{j=1}^{n_i} {\bf p}_j- \sum_{m=1}^{n_f} {\bf k}_j \right)
M \left( {\bf{p}}_1,\cdots,{\bf{p}}_{n_i} \to {\bf{k}}_1,\cdots,{\bf{k}}_{n_f} \right). 
\end{eqnarray}
By using the relation (\ref{epl}), the amplitude $|{\cal{M}}\left(E,{\bf{P}};{l} \to{l'}\right)|$ can be rewritten as 
\begin{eqnarray}
{\cal{M}}\left(E,{\bf{P}};{l} \to {l'}\right) = \int d\Pi_{n_i}({\bf{p}}_j) d\Pi_{n_f}({\bf{k}}_m) \, h_l({\bf{p}}_j) \, {\bar{h}}_{l'}({\bf{k}}_m) \, 
M\left( {\bf{p}}_1,\cdots,{\bf{p}}_{n_i} \to {\bf{k}}_1,\cdots,{\bf{k}}_{n_f}\right). 
\label{eq:amplitude}
\end{eqnarray}
The unitarity bound should be satisfied for arbitrary choices of the orthonormal basis $\{ h_l({\bf p}_j) \}$. 
This implies that scattering amplitudes have to be bounded from above  
for any normalized initial and final states, i.e. 
for any normalized functions $h_l({\bf{p}}_j)$ and $h_{l'}({\bf{k}}_j)$ 
defined on $\mathcal C_{(E,\mathbf P)}$.

\section{Unitarity Bound in Lifshitz Scalar Theory}\label{UbLs}

In this section, we derive high-energy behaviors of scattering amplitudes in the Lifshitz scalar theory.
In the following, we restrict ourselves to two-particle scattering. 
In relativistic theories, it is sufficient to consider scattering amplitudes in the center-of-mass (COM) frame
since we can use the Lorentz boost to change the reference frame.
In contrast, without the Lorentz symmetry, 
we also need to consider situations where the total momentum is non-zero. 
Such scattering processes are not related to those in the COM frame 
and may give different unitarity constraints on interaction terms. 
 
Since we are interested in high-energy scattering amplitudes 
with a large total momentum $\mathbf P$, 
we set 
\begin{eqnarray}
E \approx P^z,
\end{eqnarray}
and take the limit $P := |{\bf P}| \to \infty$. 
In this setting, 
there are two typical states which have different high energy behaviors: 
\begin{description}
\item[a)] 
Two-particle state whose momenta have the same magnitude $|{\bf p}_1| = |{\bf p}_2|$.
\item[b)]
Two-particle state with momenta ${\bf p}_1 = 0$ and ${\bf p}_2 = {\bf P}$. 
\end{description}
For \textbf{a)}, 
the energies of both particles approach infinity 
as in the case of the COM scattering. 
For \textbf{b)}, 
only one particle has high energy and the other is at rest (laboratory system), 
so that the situation is very different from the COM case. 

We construct two normalized states $|E, \mathbf P, l \rangle~(l=\alpha, \beta)$ 
which possess the properties of the two states \textbf{a)} and \textbf{b)} respectively. 
The first one, denoted by $|\alpha \rangle$, 
is the state for which the energies of two particles approach infinity as $P \rightarrow \infty$. 
It can be obtained by defining the function $h_{l=\alpha}(\mathbf p_1, \mathbf p_2)$ as
\begin{eqnarray}
h_\alpha(\mathbf p_1, \mathbf p_2) ~=~ \frac{1}{\sqrt{N_\alpha(P)}} ~ \times
\left\{
  \begin{array}{cc}
     ~1~  & \left( \mbox{$\big| |\mathbf p_1| - |\mathbf p_2| \big| \leq P/2$} \right)  \\
     ~0~  & \left( \mbox{$\big| |\mathbf p_1| - |\mathbf p_2| \big| > P/2$} \right) \\
  \end{array}
\right. .
\end{eqnarray}
The normalization factor $N_{\alpha}$ is given by
\begin{eqnarray}
N_\alpha(P) &=& \int_{I_\alpha} \frac{d^d p_1}{2 E_1}\frac{d^d p_2}{2 E_2} \,
\delta(E_1+ E_2 - E)\delta^d({\bf p}_1+{\bf p}_2-{\bf P}), 
\label{nor}
\end{eqnarray}
where the domain of integration is  
\begin{eqnarray}
I_\alpha = \left\{ ( \mathbf p_1, \mathbf p_2 ) \in \mathbb R^d \times \mathbb R^d \ \Big| \ \big| |\mathbf p_1|-|\mathbf p_2| \big| \leq P/2 \right\}.
\end{eqnarray}
By a simple dimensional analysis\footnote{
The contributions of factors in the integral are given by $d^dp_1\sim d^dp_2\propto P^d$, $E_1\sim E_2\propto P^z$, $\delta(E_1+E_2-E)\propto P^{-z}$ and $\delta^d ({\bf p}_1+{\bf p}_2-{\bf P})\propto P^{-d}$. }, 
we can estimate the asymptotic behavior of $N_\alpha$ in the limit $P \rightarrow \infty$ as 
(see Appendix \ref{appendix:integral} for more details on the phase space integrals)
\begin{eqnarray}
N_{\alpha}(P) \approx P^{d-3z}. 
\end{eqnarray}
Based on the definition of the discrete normalized state Eq.\,\eqref{epl}, 
we define the state $|\alpha \rangle$ by
\begin{eqnarray}
|\alpha \rangle ~=~ 
\frac{1}{\sqrt{N_\alpha(P)}} \int_{I_\alpha} \frac{d^d \mathbf p_1}{2 E_1} \frac{d^d \mathbf p_2}{2 E_2} \,
\delta(E_1+ E_2-E) \, \delta^d({\bf p}_1+{\bf p}_2-{\bf P}) \,
|\mathbf p_1, \mathbf p_2 \rangle. 
\end{eqnarray}

The second one, denoted by $|\beta \rangle$, is the state for which 
only one particle has high energy in the limit $P \rightarrow \infty$. 
The normalized function
\begin{eqnarray}
h_\beta(\mathbf p_1, \mathbf p_2) ~=~ \frac{1}{\sqrt{N_\beta(P)}} ~ \times
\left\{
  \begin{array}{cc}
     ~1~  & ( \mbox{for $|\mathbf p_1| \leq \epsilon$} )  \\
     ~0~  &  ( \mbox{for $|\mathbf p_1| > \epsilon$} ) \\
  \end{array}
\right.,
\end{eqnarray}
gives such a state, where $\epsilon$ is a constant satisfying $\epsilon \ll P$. 
The normalization factor $N_\beta(P)$ is defined as in Eq.\,\eqref{nor} 
with the domain of integration 
\begin{eqnarray}
I_\beta = \left\{ ( \mathbf p_1, \mathbf p_2 ) \in \mathbb R^d \times \mathbb R^d \ \big| \ |\mathbf p_1| \leq \epsilon \right\}.
\end{eqnarray}
The state $|\beta\rangle$ is defined by
\begin{eqnarray}
|\beta \rangle ~=~ 
\frac{1}{\sqrt{N_\beta(P)}} \int_{I_\beta} \frac{d^d \mathbf p_1}{2 E_1} \frac{d^d \mathbf p_2}{2 E_2} \,
\delta(E_1+ E_2-E) \, \delta^d({\bf p}_1+{\bf p}_2-{\bf P}) \,
|\mathbf p_1, \mathbf p_2 \rangle. 
\label{|beta>}
\end{eqnarray} 
Note that in the high-energy limit, this state exists only when $E = P^z + \mathcal O(P^{z-1})$.
We can estimate the asymptotic behavior of the normalization factor $N_\beta(P)$ as follows 
(see Appendix \ref{appendix:integral} for more details).  
The second particle has a large momentum $|\mathbf p_2| \approx P$, 
so that $d^d p_2$ and $E_2$ behave as $P^d$ and $P^z$, respectively. 
On the other hand, $d^d p_1$ and $E_1$ do not depend on $P$
since $\mathbf p_1$ is kept small in the high energy limit. 
After eliminating the delta function $\delta^d(\mathbf p_1 + \mathbf p_2 - {\bf P}) \approx P^{-d}$, 
the terms of order $P^z$ in the argument of $\delta(E_1+E_2-E)$ cancel with each other 
and those of order $P^{z-1}$ become dominant. 
Therefore, $\delta(E_1+E_2-E)$ is of order $P^{-(z-1)}$, 
and hence $N_{\beta}$ behaves as
\begin{eqnarray}
N_{\beta}(P) ~\approx~ P^{-2z+1},
\end{eqnarray}
in the high energy limit.

Given the two states $|\alpha \rangle$ and $|\beta \rangle$, 
we can consider the following three scattering amplitudes: 
$\mathcal M(\alpha \rightarrow \alpha)$, $\mathcal M(\beta \rightarrow \beta)$ 
and $\mathcal M(\beta \rightarrow \alpha)= \overline{\mathcal M(\alpha \rightarrow \beta)}$. 
Here, we estimate their asymptotic behavior and 
translate the unitarity bound for them into constraints on  
the scattering amplitude $M({\bf p}_1,{\bf p}_2 \to {\bf k}_1, {\bf k}_2)$ between the momentum eigenstates. 
In the estimation made below, ${\bf p}_1, {\bf p}_2$  (${\bf k}_1, {\bf k}_2$) denote the momenta of the initial (final) two particles.

\subsubsection{$\mathcal M(\alpha \rightarrow \alpha)$}
In this scattering process, 
all the momenta of the initial and the final particles approach infinity, i.e. $|{\bf p}_1|,|{\bf p}_2|,  |{\bf k}_1|, |{\bf k}_2| \approx P$. 
In the integral form of the scattering amplitude Eq.\,\eqref{eq:amplitude}, 
both the initial and the final phase space factors $d \Pi({\bf p})$ and $d \Pi({\bf k})$ 
have the same high-energy behavior as $N_{\alpha}$.
Assuming that the leading order behavior of $M ({\bf p}_1,{\bf p}_2 \to {\bf k}_1, {\bf k}_2)$ 
on the support of $h_\alpha$ is $M ({\bf p}_1,{\bf p}_2 \to {\bf k}_1, {\bf k}_2) \approx P^a$, 
we can estimate the amplitude as
\begin{eqnarray}
\mathcal M(\alpha \rightarrow \alpha) \approx P^{a-3z+d}. 
\end{eqnarray}
Therefore, the unitarity bound $\mathcal M(\alpha \rightarrow \alpha) \leq 1$ is satisfied if
\begin{eqnarray}
M ({\bf p}_1,{\bf p}_2 \to {\bf k}_1, {\bf k}_2) \approx P^a ~~~ \mbox{with ~ $a \le 3z-d$}. 
\label{bb}
\end{eqnarray}

\subsubsection{$\mathcal M(\beta \rightarrow \beta)$}

In this case, the initial and the final momenta of the first particle are small $|{\bf p}_1|,  |{\bf k}_1| \propto P^0$ 
and those of the second particle become large $|{\bf p}_2|, |{\bf k}_2| \propto P$ for large $P$. 
Both $d \Pi({\bf p})$ and $d \Pi({\bf k})$ have the same high-engnergy behavior as $N_{\beta}$.
Therefore, assuming that $M({\bf p}_1,{\bf p}_2 \to {\bf k}_1, {\bf k}_2) \approx P^a$, we find that 
\begin{eqnarray}
\mathcal M(\beta \rightarrow \beta) \approx P^{a-2z+1}. 
\end{eqnarray}
The unitarity bound $\mathcal M(\beta \rightarrow \beta) \leq 1$ is satisfied if
\begin{eqnarray}
M({\bf p}_1,{\bf p}_2 \to {\bf k}_1, {\bf k}_2) \approx P^a ~~~ \mbox{with ~ $a \le 2z-1$}. 
\label{aa} 
\end{eqnarray}

\subsubsection{$\mathcal M(\beta \rightarrow \alpha)$}

For the scattering from $|\beta \rangle$ to $|\alpha \rangle$, 
only the initial momentum of the first particle is small $|{\bf p}_1| \propto P^0$ and 
the others are large $|{\bf p}_2|, |{\bf k}_1|,  |{\bf k}_2| \propto P$.
The measures $d \Pi({\bf p})$ and $d \Pi({\bf k})$ have the same high-energy behaviors as $N_{\beta}$ and $N_{\alpha}$, respectively.
Assuming that $M ({\bf p}_1,{\bf p}_2 \to {\bf k}_1, {\bf k}_2) \approx P^a$, we find that 
\begin{eqnarray}
\mathcal M(\beta \rightarrow \alpha) \approx P^{a - \frac{5z-d-1}{2}}.
\end{eqnarray}
The unitarity bound $\mathcal M(\beta \rightarrow \alpha) \leq 1$ is satisfied if
\begin{eqnarray}
M({\bf p}_1,{\bf p}_2 \to {\bf k}_1, {\bf k}_2) \approx P^a ~~~ \mbox{with ~ $a \le (5z-d-1)/2$}. 
\label{ab}
\end{eqnarray}

\section{Constraints for Vertices from Unitarity Bound}\label{Sec:const}

In this section, 
we see that the unitarity conditions derived in Sec.\,\ref{UbLs} agree with 
the extended (and conventional) PCR conditions for quartic and cubic interactions. 

\subsection{Quartic Interactions}\label{QI}

\begin{figure}[tb]
\centering
\includegraphics[width=60mm]{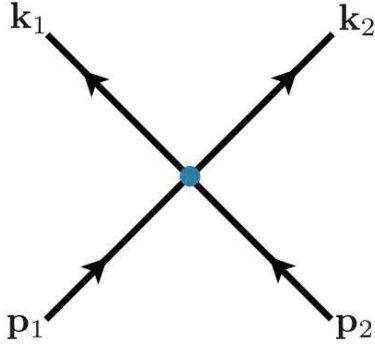}
\caption{four point tree-level diagram}
\label{fig:4point}
\end{figure}

We first consider the following generic quartic interaction term: 
\begin{eqnarray}
S_{4}=\int dt d^dx \left(\partial_x^{a_1} \phi\right) \left(\partial_x^{a_2} \phi \right) \left(\partial_x^{a_3} \phi\right) \left(\partial_x^{a_4} \phi\right).
\end{eqnarray}
Without loss of generality, we can set $0 \leq a_1\leq a_2 \leq a_3 \leq a_4$. 
The momentum dependence of scattering amplitude (Fig.\,\ref{fig:4point}) is given by
\begin{eqnarray}
M(\mathbf p_1, \mathbf p_2 \rightarrow \mathbf k_1, \mathbf k_2) ~\propto~ |\mathbf p_1|^{a_1} |\mathbf p_2|^{a_2} |\mathbf k_1|^{a_3} 
|\mathbf k_2|^{a_4} + (\mbox{permutations w.r.t. $a_i$}),
\end{eqnarray}
where $\mathbf p_1$, $\mathbf p_2$, $\mathbf k_1$ and $\mathbf k_2$ are the momentum of external lines.
The extended PCR conditions \eqref{PCR} and \eqref{eq:ePCR} require that
\begin{eqnarray}
a_1 + a_2 + a_3 + a_4 &\le& \hspace{3mm} 3z-d,  \label{eq:e=1} \\
a_2 + a_3 + a_4 \hspace{4mm} &<& (5z-d)/2, \hspace{5mm} \Rightarrow\hspace{5mm} a_2 + a_3 + a_4  \le (5z-d-1)/2 \label{eq:e=2} \\
a_3 + a_4 \hspace{9mm} &<& \hspace{7mm} 2 z, \hspace{13mm} \Rightarrow \hspace{9mm} a_3 + a_4 \hspace{5mm} \le \hspace{5mm} 2z-1,\label{eq:e=3}
\end{eqnarray}
where we have rewritten the inequalities with $<$ into those with $\le$ 
by using the fact that $a_1$, $a_2$, $a_3$, $a_4$, $d$, and $z$ are all integers. 
In the following, we show that these conditions agree with the tree-level unitarity conditions. 

\subsubsection{$|\alpha \rangle \rightarrow |\alpha \rangle$}
For the scattering from $|\alpha \rangle$ to $|\alpha \rangle$, 
the momenta of all the external lines are proportional to $P$ for large $P$, 
so that the high energy behavior of $M(\mathbf p_1, \mathbf p_2 \rightarrow \mathbf k_1, \mathbf k_2)$ is given by
\begin{eqnarray}
M(\mathbf p_1, \mathbf p_2 \rightarrow \mathbf k_1, \mathbf k_2) ~\approx~ P^{a_1 + a_2 + a_3 + a_4}. 
\end{eqnarray}
Thus, the condition (\ref{bb}) gives the constraint 
\begin{eqnarray}
a_1+ a_2 + a_3 + a_4 \leq 3z-d.
\end{eqnarray}
This coincides with the conventional PCR condition (\ref{eq:e=1}). 

\subsubsection{$|\beta \rangle \rightarrow |\beta \rangle$}
For the scattering from $|\beta \rangle$ to $|\beta \rangle$, 
two external momenta are proportional to $P$, 
so that the leading term in $M(\mathbf p_1, \mathbf p_2 \rightarrow \mathbf k_1, \mathbf k_2)$ is given by
\begin{eqnarray}
M(\mathbf p_1, \mathbf p_2 \rightarrow \mathbf k_1, \mathbf k_2) ~\approx~ P^{a_3 + a_4}. 
\end{eqnarray}
Thus, the condition \eqref{aa} gives
\begin{eqnarray}
a_3 + a_4 \leq 2z-1. 
\end{eqnarray}
This agrees with the \underline{extended} PCR condition (\ref{eq:e=3}). 

\subsubsection{$|\beta \rangle \rightarrow |\alpha \rangle$}
For the scattering from $|\beta \rangle$ to $|\alpha \rangle$, 
three external momenta approach infinity, and thus
\begin{eqnarray}
M(\mathbf p_1, \mathbf p_2 \rightarrow \mathbf k_1, \mathbf k_2) ~\approx~ P^{a_2 + a_3 + a_4}.
\end{eqnarray}
The constraint \eqref{ab} gives  
\begin{eqnarray}
a_2+a_3+a_4 \le (5z-d-1)/2.
\end{eqnarray}
This is equivalent to the \underline{extended} PCR condition (\ref{eq:e=2}).

\subsection{Cubic Interactions}\label{CI}

Next, we derive the constraints for cubic interactions from the unitarity conditions. 
The generic form of cubic interactions is given by 
\begin{eqnarray}
S_{3} = \int dt d^dx \left(\partial_x^{a_1} \phi\right)\left(\partial_x^{a_2} \phi\right)\left(\partial_x^{a_3} \phi\right), 
\label{eq:3-vertex}
\end{eqnarray}
with $0 \leq a_1\le a_2\leq a_3$. 
The extended PCR condition implies that
\begin{eqnarray}
a_1 + a_2 + a_3 &\le& (5z-d)/2 , \label{eq:PCR_cubic} \\ 
a_2 + a_3 &<& ~~~~ 2z \hspace{15mm} \Rightarrow \hspace{5mm} a_2 + a_3 \le 2z-1,
\label{eq:ePCR_cubic}
\end{eqnarray}
where we have rewritten the second inequality by using the fact that $a_1$, $a_2$, $a_3$, $d$ and $z$ are integers. 
For the interaction term \eqref{eq:3-vertex}, 
the contribution of the corresponding 3-point vertex takes the form
\begin{eqnarray}
V(\mathbf p_1, \mathbf p_2, \mathbf p_1 + \mathbf p_2) ~=~ |\mathbf p_1|^{a_1} |\mathbf p_2|^{a_2} |\mathbf p_1 + \mathbf p_2|^{a_3} + (\mbox{permutations w.r.t. $a_i$}). 
\end{eqnarray}
\begin{figure}[tb]
\begin{minipage}{0.49\hsize}
\begin{center}
\includegraphics[width=40mm]{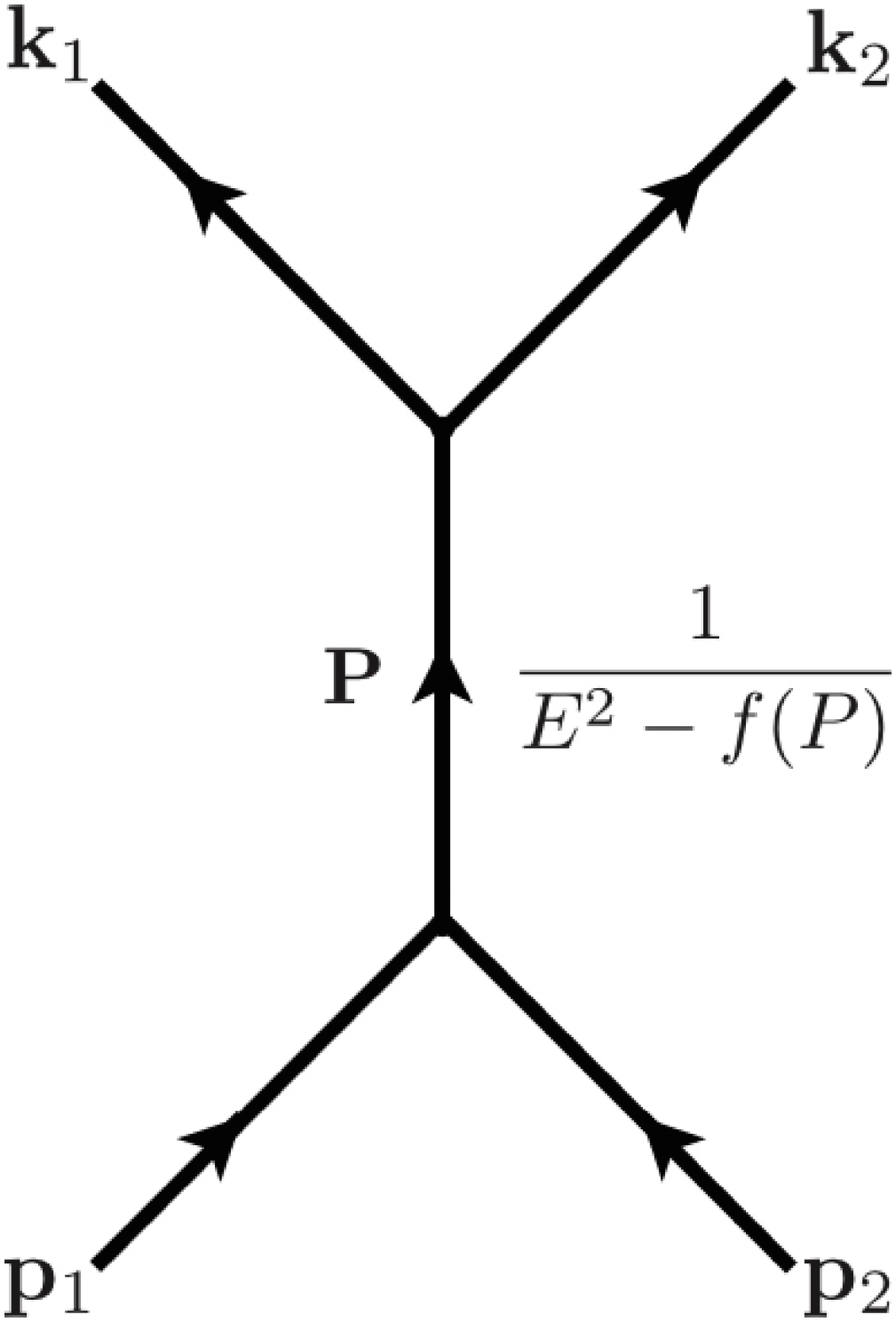}
\caption{s-channel diagram}
\label{fig:s-channel}
\end{center}
\end{minipage}
\begin{minipage}{0.49\hsize}
\begin{center}
\includegraphics[width=60mm, bb=00 -50  575 423]{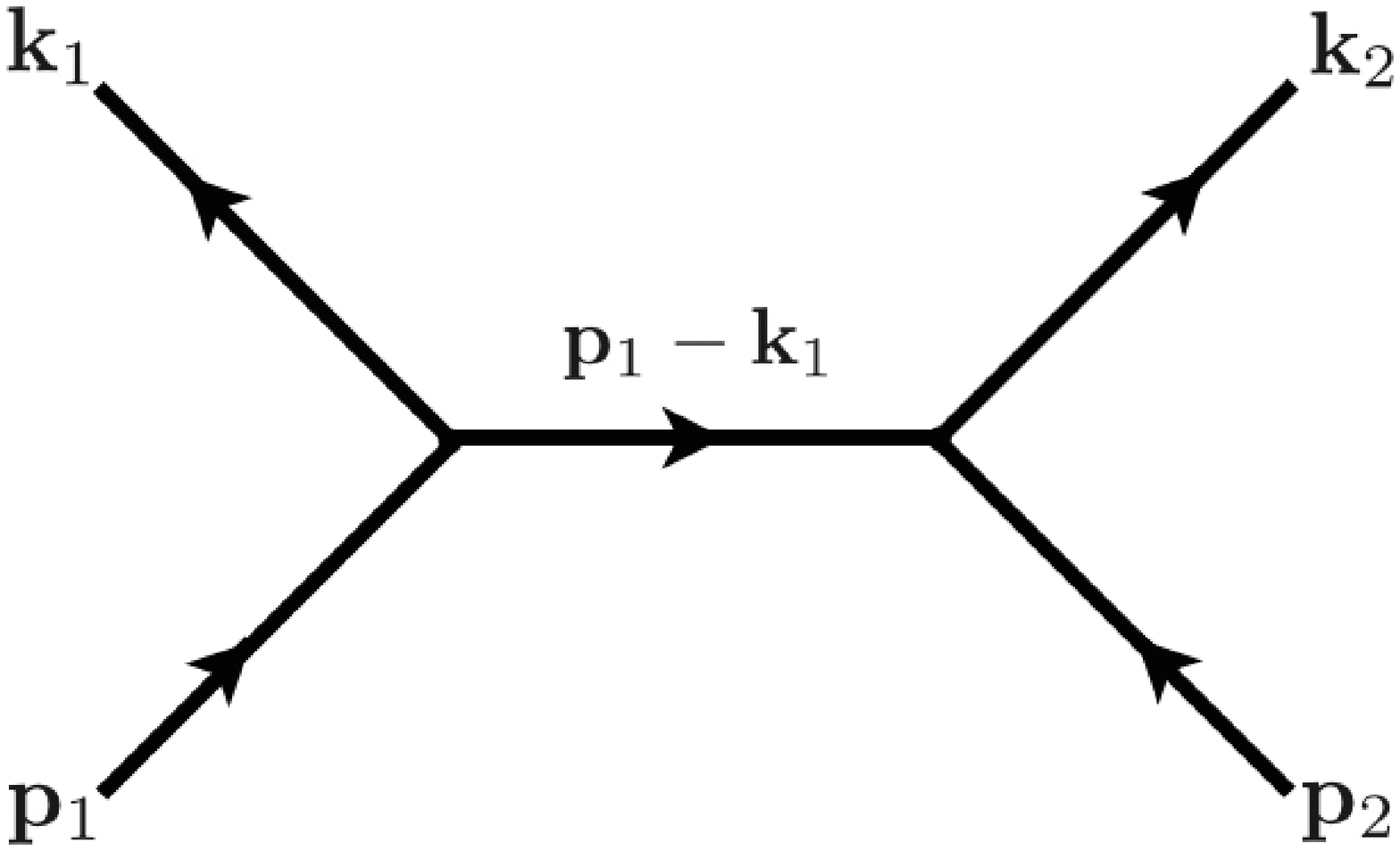}
\caption{t-channel diagram}
\label{fig:t-channel}
\end{center}
\end{minipage}
\end{figure}

\subsubsection{s-channel}
Let us first consider the s-channel scattering amplitudes (Fig.\,\ref{fig:s-channel}): 
\begin{eqnarray}
M(\mathbf p_1, \mathbf p_2 \rightarrow \mathbf k_1, \mathbf k_2) ~\propto~
\frac{V(\mathbf p_1, \mathbf p_2, \mathbf p_1 + \mathbf p_2) V(\mathbf k_1, \mathbf k_2,\mathbf k_1 + \mathbf k_2)}{E^2 - f(P^2)}.
\end{eqnarray}
The denominator $E^2 - f(P^2)$ can be roughly estimated as $P^{2z}$ in the UV limit.\footnote{
This estimation seems invalid for $E = P^z + \mathcal O(P^{z-1})$. 
Such a situation always occurs if the initial and/or final states are $| \beta \rangle$. 
However, as we will see in the next section, 
the propagator can have contributions from the imaginary part of the self-energy diagrams ${\rm Im} \, \Sigma(E,P)$. 
From a simple dimensional analysis, 
we can find that marginal interactions can give 
the leading contribution to the imaginary part of the self-energy 
${\rm Im} \, \Sigma(E,P) \approx P^{2z}$. }

When all the external momenta approach infinity ($|\alpha \rangle \rightarrow |\alpha \rangle$ scattering), 
the scattering amplitude can be estimated as 
\begin{eqnarray}
M(\mathbf p_1, \mathbf p_2 \rightarrow \mathbf k_1, \mathbf k_2) \approx P^{2a_1+2a_2+2a_3-2z}.
\end{eqnarray}
From the condition (\ref{bb}), we find that 
\begin{eqnarray}
a_1+a_2+a_3 \le (5z-d)/2. \label{3-1}
\end{eqnarray}
For the scattering from $|\beta \rangle$ to $|\beta \rangle$, 
the leading part of the scattering amplitude is given by
\begin{eqnarray}
M(\mathbf p_1, \mathbf p_2 \rightarrow \mathbf k_1, \mathbf k_2) \approx P^{2a_2+2a_3-2z}. 
\end{eqnarray}
The condition (\ref{aa}) implies that
\begin{eqnarray}
a_2 + a_3 \le 2z-\frac{1}{2} \hspace{5mm} \Rightarrow \hspace{5mm} a_2 + a_3 \le 2z-1, \label{3-2}
\end{eqnarray}
where we have used the fact that all $a_i$ and $z$ are integers.
These two conditions are equivalent to the extended PCR conditions \eqref{eq:PCR_cubic} and \eqref{eq:ePCR_cubic}. 

For the scattering from $|\beta \rangle$ to $|\alpha \rangle$, 
the leading order behavior of the amplitude becomes
\begin{eqnarray}
M(\mathbf p_1, \mathbf p_2 \rightarrow \mathbf k_1, \mathbf k_2) \approx P^{a_1+2a_2+2a_3-2z}. 
\end{eqnarray}
The condition (\ref{ab}) implies that
\begin{eqnarray}
a_1+2 a_2+2 a_3 \le (9z-d-1)/2. \label{3-3}
\end{eqnarray}
This constraint is automatically satisfied if both Eqs.\,\eqref{3-1} and \eqref{3-2} are met. 

\subsubsection{t-(and u-)channels}
Next, we consider the t-(and u-)channel scattering amplitude (Fig.\,\ref{fig:t-channel}).
In most cases, the magnitude of internal momentum $|\mathbf p_1 + \mathbf p_2|$ becomes of order $\mathcal O(P)$ 
and the estimation is the same as that in the case of the s-channel.
With a fine tuning of the initial and final states,
the internal momentum can be of order $\mathcal O(P^0)$ in the limit $P \rightarrow \infty$. 
We show that such scattering amplitudes have different asymptotic behaviors 
but give no additional constraint for the cubic interactions. 

For the scattering from $|\alpha \rangle$ to $|\alpha \rangle$, 
the leading behavior of the scattering amplitude can be estimated as 
\begin{eqnarray}
M(\mathbf p_1, \mathbf p_2 \rightarrow \mathbf k_1, \mathbf k_2)\approx P^{2a_2+2a_3}.
\end{eqnarray}
The condition (\ref{bb}) implies that
\begin{eqnarray}
a_2+a_3 \le (3z-d)/2.
\end{eqnarray}
Since the right hand side is smaller than or equal to $(2z-1)$, 
this condition is weaker than the inequality (\ref{3-2}). 
For the scattering from $|\beta \rangle$ to $|\beta \rangle$, 
the leading behavior of the scattering amplitude is given by
\begin{eqnarray}
M(\mathbf p_1, \mathbf p_2 \rightarrow \mathbf k_1, \mathbf k_2) \approx P^{a_2+a_3}.
\end{eqnarray}
From the condition (\ref{aa}), we obtain the constraint
\begin{eqnarray}
a_2+a_3 \le 2z- 1. 
\end{eqnarray}
This is the same condition as the inequality (\ref{3-2}).
If momenta of three external lines approach infinity, 
the momentum of internal line cannot be finite.
In summary, these types of scattering amplitudes gives no additional constraint for the cubic interactions. 

\section{On-shell Intermediate States and Resonances}\label{SecTUOS}
In Lifshitz-type theories, there exists a peculiar phenomenon 
in scattering processes which is absent in relativistic theories,   
namely the appearance of on-shell intermediate states.  
When the intermediate particle has an on-shell momentum, 
the high-energy s-channel amplitude appears to violate the unitarity bound  
from the naive estimate in the previous section. 
However, in such cases, the tree-level discussion breaks down and 
we should correctly take into account the corresponding resonance effects.
In this section, we show that the unitarity is not violated 
if all the extended PCR conditions are satisfied. 

\subsection{Resonances Induced by Identical Particles}
\begin{figure}[tb]
\begin{minipage}{0.45\hsize}
\begin{center}
\includegraphics[width=80mm]{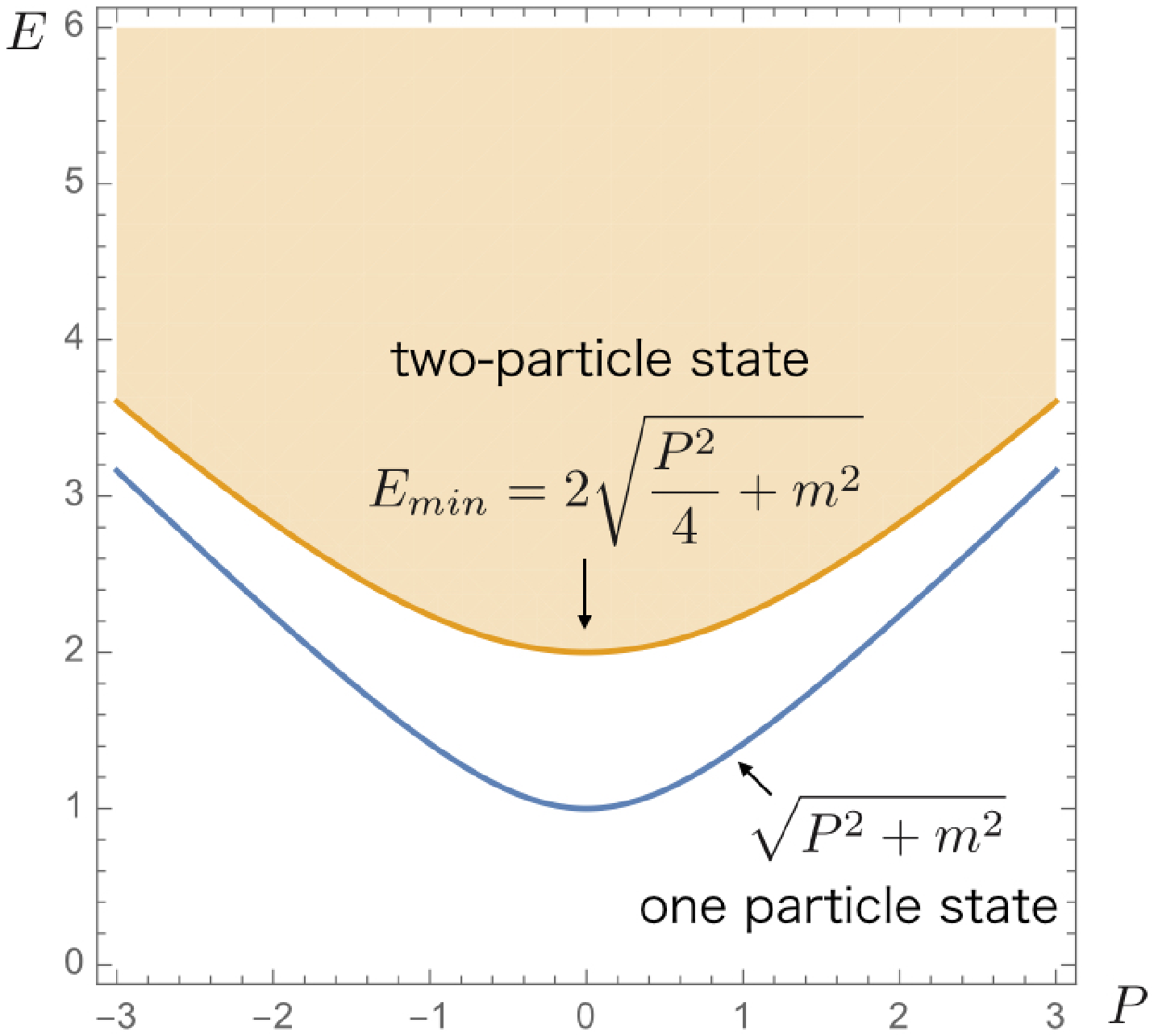}
\caption{Spectrum for z=1 and $f(p^2) = p^2 + m^2$}
\label{fig:Lorentz}
\end{center}
\end{minipage}
\hspace{10mm}
\begin{minipage}{0.45\hsize}
\begin{center}
\includegraphics[width=80mm]{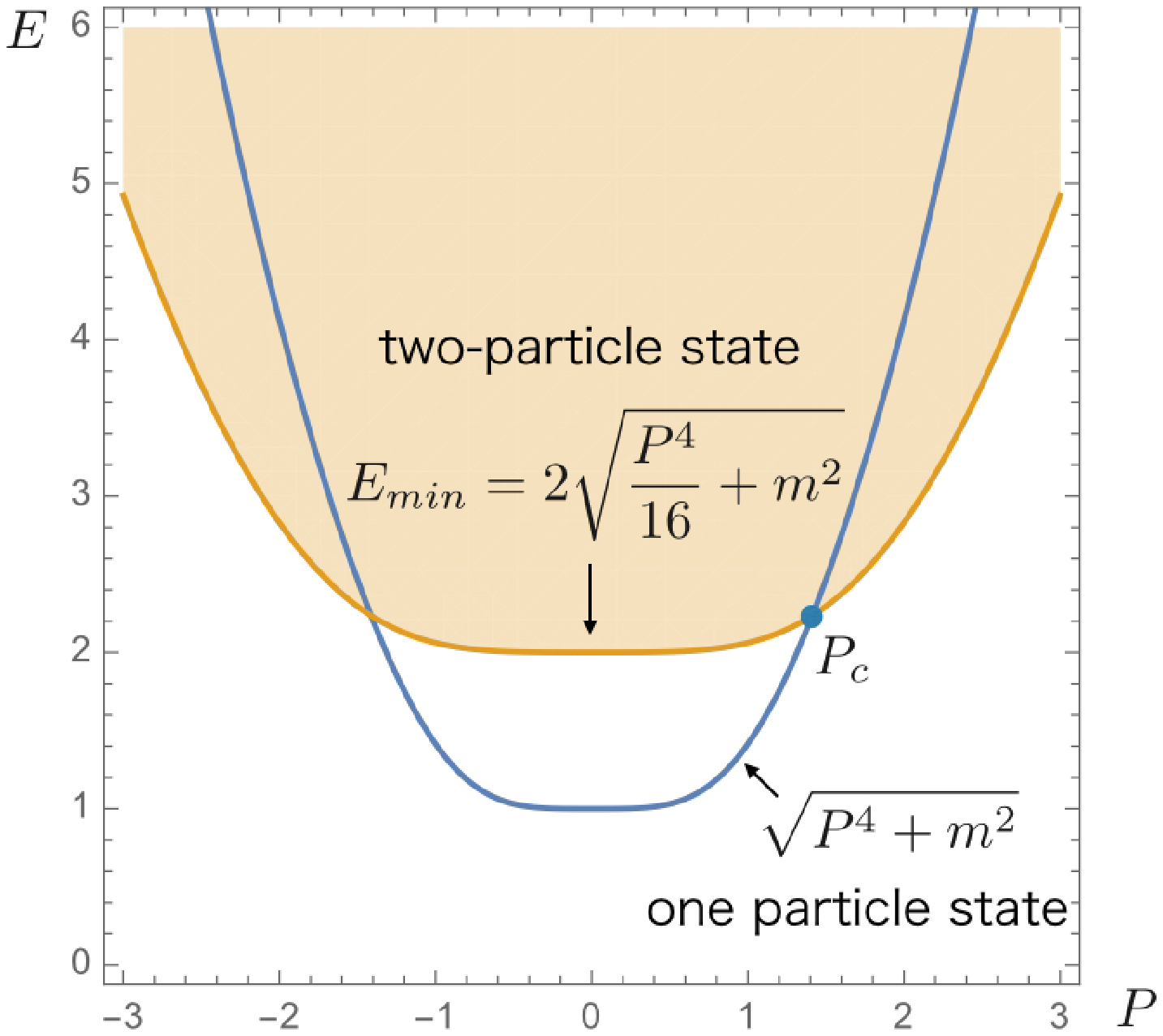}
\caption{Spectrum for z=2 and $f(p^2) = p^4 + m^2$}
\label{fig:Lifshitz}
\end{center}
\end{minipage}
\end{figure}

In relativistic theories, any single particle cannot decay into two (or more) identical particles. 
In contrast, in Lifshitz-type theories, 
single particle states can decay into multi-particle states  
if its momentum is larger than a certain stability threshold. 
This can be seen from the dispersion relation given in Eq.\,\eqref{DR}.
For two-particle states, the total energy and momentum satisfy the following inequality:
\begin{eqnarray}
E_{two} \geq E_{min} := 2 \sqrt{f\bigl( P^2/4\bigr)},
\end{eqnarray}
where the equality holds only when two particles have the same momentum, i.e.
\begin{eqnarray}
{\bf p}_{1}= {\bf p}_{2}=\frac{1}{2}{\bf P}.
\end{eqnarray}
A single particle can decay into two particles, 
if its energy $E_{one}$ is larger than $E_{min}$. 
In the Lifshitz scaling theory with $z>1$, the difference $E_{min}-E_{one}$ for $P=0$ has the positive value $\sqrt{f(0)}$, 
whereas it becomes negative for large $P$ (see Fig.\,\ref{fig:Lorentz} and Fig.\,\ref{fig:Lifshitz}): 
\begin{eqnarray}
E_{min} - E_{one} = \left( \frac{1}{2^{z-1}}-1\right) P^z + {\cal{O}}(P^{z-1}) < 0.
\end{eqnarray}
Therefore, there exists a stability threshold $P_c$ at which $E_{min} - E_{one} = 0$. 
This also implies that on-shell intermediate states can appear in the s-channel scattering processes 
if the total momentum is greater than $P_c$. 

When there exists an on-shell intermediate state, 
we need to take into account the effect of the resonance\footnote{
In theories with the Lorentz symmetry, 
resonance occurs only at a finite Mandelstam variable $s$, 
so that we do not need to consider the resonance 
in the discussion of the unitarity bound for high energy scattering amplitudes. 
}.
In terms of the tree-level propagator $G_0$ and the 1PI self-energy $-i\Sigma(E, P)$, 
the full propagator $G$ is given by 
\begin{eqnarray}
G &=& G_0 + G_0 (-i\Sigma) G_0 + G_0 (-i\Sigma) G_0 (-i\Sigma) G_0 + \cdots ~=~ \frac{i}{E^2-f(P^2)-\Sigma} \ ,
\end{eqnarray}
where we have used the tree-level propagator derived from the quadratic action (\ref{2nd});
\begin{eqnarray}
G_0= \frac{i}{E^2-f(P^2)}.
\end{eqnarray}
Imposing the on-shell renormalization condition
\begin{eqnarray}
{\rm {Re}}\hspace{1mm}{ \Sigma(E,P) \left|_{E=\sqrt{f(P^2)}} \right. =0},
\end{eqnarray}
we find that the propagator at $E=\sqrt{f(P^2)}$ is given by the imaginary part of the self-energy $\Sigma$:
\begin{eqnarray}
\left.G \left(E=\sqrt{f(P^2)}, P\right) ~=~ -\frac{1}{{\rm {Im}}{\hspace{1mm}}\Sigma(E, P)} \right|_{E=\sqrt{f(P^2)}}.
\end{eqnarray}
Therefore, the propagator does not diverge at $E=\sqrt{f(P^2)}$.

\subsection{s-channel Amplitudes with On-shell Internal States}

Let us estimate the s-channel amplitudes with the on-shell internal propagator 
and see that the unitarity condition is satisfied if all conditions shown in Sec.\,\ref{UbLs} are obeyed.

\begin{figure}[tb]
\begin{center}
\vspace{-10mm}
\includegraphics[width=90mm]{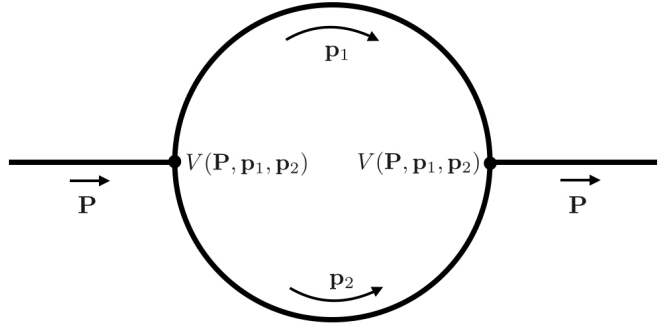}
\caption{Self energy diagram}
\label{fig:selfenergy}
\end{center}
\end{figure}
\begin{figure}[tb]
\begin{center}
\includegraphics[width=110mm]{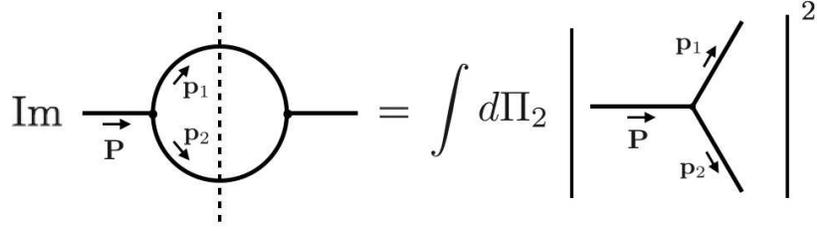}
\caption{Cutting rule}
\label{fig:cutting}
\end{center}
\end{figure}

The momentum dependence of the self-energy can be estimated as follows.
The leading order contribution to the 1PI self energy $-i\Sigma(E, P)$ is give by 
the one-loop diagram (see Fig.\,\ref{fig:selfenergy}), which takes the form
\begin{eqnarray}
-i\Sigma_{\text{1-loop}}(E, P)= i \int d\omega d^dp \left| V({\bf P},{\bf p}_1,{\bf p}_2) \right|^2 G_1 G_2,
\end{eqnarray}
with ${\bf p_1}:={\bf P}/2 + {\bf p}$, ${\bf p}_2 := {\bf P}/2 -{\bf p}$ and 
\begin{eqnarray}
G_1= \frac{1}{(E/2 + \omega)^2 - f(p_1^2) + i \epsilon}, \qquad  G_2= \frac{1}{(E/2 + \omega)^2 - f(p_2^2) + i \epsilon}.
\end{eqnarray}
Here, $V({\bf P},{\bf p}_1,{\bf p}_2)$ is a three-point vertex, 
for which we use the generic form (\ref{eq:3-vertex}) for definiteness.
To evaluate the imaginary part of $\Sigma$, 
we use the following relation:
\begin{eqnarray}
\frac{1}{x+i\epsilon}= {\cal P} \frac{1}{x} -i\pi \delta (x),
\end{eqnarray}
where ${\cal P} \frac{1}{x}$ denotes the principal value of $1/x$. 
Then, the imaginary part of $\Sigma$ can be rewritten as 
\begin{eqnarray}
&&{\rm {Im}} \hspace{1mm} \Sigma(E, P) \nonumber \\
&&\qquad
\sim \pi^2  \int d\omega d^dp \left| V({\bf P},{\bf p}_1,{\bf p}_2)  \right|^2 \delta \left((E/2 + \omega)^2 - f(p_1^2) \right) \delta\left((E/2 + \omega)^2 - f(p_2^2) \right) \nonumber \\
&&\qquad
=\pi^2  \int \frac{d^dp_1}{2 E_1} \frac{d^dp_2}{2 E_2} \left| V({\bf P},{\bf p}_1,{\bf p}_2) \right|^2 
\delta \left( E_1+E_2-E  \right) \delta^d \left( {\bf p}_1+{\bf p}_2- {\bf P}\right),
\end{eqnarray}
where we have used the fact that the product of the principal values do not contribute the imaginary part. 
Note that this is the cutting rule which relates 
the imaginary part of the one-loop self energy diagram and the tree-level decay diagram (see Fig.\,\ref{fig:cutting}).
Suppose that the interaction term (\ref{eq:3-vertex}) gives the leading order contribution among cubic order interactions. 
Then, ${\rm {Im}} \hspace{1mm} \Sigma(E, P)$ can be written as
\begin{eqnarray}
{\rm {Im}}\hspace{1mm} \Sigma(E, P) ~\sim~ 
\pi^2  \int d \Pi_2(\mathbf p) \, \Big| p_1^{a_1}p_2^{a_2} |{\bf p}_1+{\bf p}_2|^{a_3} + (\mbox{permutations w.r.t. }a_i) \Big|^2.
\end{eqnarray}
There are two regions in the integration domain either of which can give 
a leading contribution to ${\rm {Im}}\hspace{1mm} \Sigma$: 
\begin{enumerate}
\item
The region where both $p_1$ and $p_2$ are of order $\mathcal O(P)$, 
\item
The region where only $p_1$ (or $p_2$) is of order $\mathcal O(P)$.
\end{enumerate}
The contributions from region 1 and 2 are respectively estimated as $P^{\gamma_1}$ and $P^{\gamma_2}$
with
\begin{eqnarray}
\gamma_1 := d-3z+2(a_1+a_2+a_3), ~~~~~ \gamma_2:={-2z+1+2(a_2+a_3)},
\end{eqnarray}
where the estimation of the delta function for the latter case is the same as that in the discussion below Eq.\,(\ref{|beta>}).
Therefore, the asymptotic behavior of ${\rm {Im}} \, \Sigma$ is given by
\begin{eqnarray}
{\rm {Im}} \, \Sigma \approx P^{{\rm max}(\gamma_1, \gamma_2)}.
\end{eqnarray}

Let us see the s-channel diagram Fig.\,\ref{fig:s-channel} with an on-shell intermediate state. 
Since the interaction term Eq.\,(\ref{eq:3-vertex}) gives the leading contribution by assumption, 
diagrams constructed with the other 3-point vertices are sub-leading.
Therefore, it is sufficient to check the unitarity for the s-channel amplitudes 
constructed with the vertices corresponding to Eq.\,(\ref{eq:3-vertex}).

\subsubsection{$|\alpha \rangle \rightarrow |\alpha \rangle$}

For $\gamma_1\ge \gamma_2$, 
the asymptotic behavior of $M(\mathbf p_1, \mathbf p_2 \rightarrow \mathbf k_1, \mathbf k_2)$ is given by
\begin{eqnarray}
M(\mathbf p_1, \mathbf p_2 \rightarrow \mathbf k_1, \mathbf k_2) ~\approx~ P^{3z-d}. 
\end{eqnarray}
This satisfies the unitarity condition (\ref{bb}).
For $\gamma_1\le\gamma_2$, 
the amplitude $M(\mathbf p_1, \mathbf p_2 \rightarrow \mathbf k_1, \mathbf k_2)$ can be estimated as
\begin{eqnarray}
M(\mathbf p_1, \mathbf p_2 \rightarrow \mathbf k_1, \mathbf k_2) ~\approx~ P^{2a_1-1+2z}.
\end{eqnarray}
Since the assumption $\gamma_1\le \gamma_2$ implies that ${2a_1-1+2z} \le 3z-d$, 
the unitarity condition (\ref{bb}) is satisfied.

\subsubsection{$|\beta \rangle \rightarrow |\beta \rangle$}

For $\gamma_1\ge \gamma_2$, 
the leading term in $M(\mathbf p_1, \mathbf p_2 \rightarrow \mathbf k_1, \mathbf k_2)$ is given by
\begin{eqnarray}
M(\mathbf p_1, \mathbf p_2 \rightarrow \mathbf k_1, \mathbf k_2) ~\approx~ P^{-d+3z-2a_1}.
\end{eqnarray}
Since $-d+3z-2a_1\leq 2z-1$ for $\gamma_1\ge \gamma_2$, 
the unitarity condition \eqref{aa} is satisfied.
For $\gamma_1\le\gamma_2$, we can estimate $M(\mathbf p_1, \mathbf p_2 \rightarrow \mathbf k_1, \mathbf k_2)$ as
\begin{eqnarray}
M(\mathbf p_1, \mathbf p_2 \rightarrow \mathbf k_1, \mathbf k_2) ~\approx~ P^{2z-1}. 
\end{eqnarray}
This amplitude satisfies the unitarity condition \eqref{aa}.

\subsubsection{$|\alpha \rangle \rightarrow |\beta \rangle$}
For $\gamma_1\ge \gamma_2$, the leading behavior of the amplitude can be estimated as
\begin{eqnarray}
M(\mathbf p_1, \mathbf p_2 \rightarrow \mathbf k_1, \mathbf k_2) ~\approx~ P^{-d+3z-a_1}. 
\end{eqnarray}
Since the inequality $\gamma_1\ge \gamma_2$ gives ${-d+3z-a_1}\leq \frac{5z-d-1}{2}$,  
this satisfies the unitarity condition \eqref{ab}. 
For $\gamma_1\le \gamma_2$, 
the leading term of $M(\mathbf p_1, \mathbf p_2 \rightarrow \mathbf k_1, \mathbf k_2)$ reads
\begin{eqnarray}
M(\mathbf p_1, \mathbf p_2 \rightarrow \mathbf k_1, \mathbf k_2) ~\approx~ P^{a_1+2z-1}. 
\end{eqnarray}
Since the inequality $\gamma_1\le \gamma_2$ can be rewitten as
$a_1-1+2z \leq \frac{5z-d-1}{2}$, the condition \eqref{ab} is satisfied.

\section{Summary} \label{summary}

In this paper, we have studied the renormalizability and unitarity of the self-interacting Lifshitz scalar theory. 
Because of the anisotropic scaling between the space and time directions, 
the conditions for the renormalizability and unitarity are different from those in the relativistic theories. 
We have derived the conditions for the cubic and quartic interaction terms 
and shown that the renormalizability and unitarity require equivalent conditions. 
This is one of the evidences for the equivalence between the renormalizability and unitarity in generic field theories. 

We have shown in Sec.\,\ref{Lfr} that 
the condition for the renormalizability is extended 
from the conventional one which we usually apply to the relativistic theory. 
Because of the anisotropic scaling, the dimension of the Lifshitz scalar field can be negative. 
In such cases, even if an interaction term satisfies the conventional PCR condition \eqref{PCR}, 
a part of the term can have a large dimension which excesses the limit for a renormalizable interaction. 
If only the part of the operator contributes to a loop integral, 
it behaves like a non-renormalizable term in the sense of the conventional PCR, 
i.e. it increases the degree of divergence of the loop integral. 
Therefore, such a type of interaction term is non-renormalizable. 
We have derived the constraints that rule out this type of interaction term. 
We call these conditions the extended PCR conditions. 
Their explicit forms are given as inequalities for the dimensions of coupling constants (\ref{pcr}) and (\ref{eq:renormalization_condition}) 
(or for the dimensions of interaction terms (\ref{PCR}) and (\ref{eq:ePCR})). 

We have also derived the conditions for cubic and quartic interactions 
from the tree-level unitarity bound for two-particle scattering amplitudes. 
Unlike in the relativistic case, scattering amplitudes have the reference frame dependence, 
a variety of conditions are required for unitarity. 
We have seen in Sec.\,\ref{Sec:const} that the conditions for unitarity are equivalent to the extended PCR conditions. 
We have also investigated, in Sec.\,\ref{SecTUOS}, the cases where the intermediate propagator in the s-channel diagram satisfies the on-shell condition. 
The unitarity bound for the s-channel diagram with the on-shell intermediate propagator appears to give stronger conditions, 
but actually the tree-level argument is not applicable in such situations.
We have shown in Sec.\,\ref{SecTUOS} that if the nonlinear effect, that is resonance, is taken into account, 
interaction terms satisfying the extended PCR conditions do not violate the unitarity bound. 

From our results, the equivalence between renormalizability and unitarity 
is expected to hold true even in non-relativistic theories. 
This relation is helpful to investigate the renormalizability of the HL gravity.  
Moreover, the extended PCR condition gives constraints on the form of the action for matter fields in the HL gravity. 
To preserve the anisotropic scaling property of the HL gravity, 
matter fields are also expected to follow the same anisotropic scaling. 
The dimensions of fields are zero for $d=z=3$, and the fields have to satisfy the extended PCR.
Meanwhile, for the gravitational action in the HL gravity with $d=z=3$, 
the conventional PCR can still work because of the symmetry. 
The invariance under the foliation preserving diffeomorphism constrains the form of the possible gravitational operators, 
and all of their dimensions are positive in the HL gravity with $d=z=3$. 
Without non-positive dimensional operator, the conventional PCR is enough for the renormalization. 
It may be possible to see the cancellation among the bad-behaved diagrams 
both for renormalizability and for unitarity in the same way as the Weinberg-Salam theory. 
In the HL gravity with $d=3$ and $z \ge 7$, the three dimensional curvature has non-positive dimension, 
and thus the extended PCR would be required for renormalization. 
To investigate the relevance of symmetries to renormalizability and unitarity, 
it is important to extend our study to Lifshitz-type theories with symmetries 
such as non-linear sigma models \cite{Das:2009ba, Anagnostopoulos:2010gw, Gomes:2013jba, Fujimori:2015lda}.

\section*{Acknowledgement}

The work of T. F. is supported in part by the Japan Society for the Promotion of Science (JSPS) Grantin-Aid for Scientific Research (KAKENHI Grant No. 25400268) and by the MEXT-Supported Program for the Strategic Research Foundation at Private Universities"Topological Science"(Grant No. S1511006).The work of K. I. is funded by FPA2013-46570-C2-2-P, AGAUR 2009-SGR-168, and MDM-2014-0369 of ICCUB (Unidad de Excelencia 'Maria de Maeztu'). T. K. benefitted much from his visits to NTU. He wishes to thank Pei-Ming Ho for the kind support for the visit. He also thanks Kei-ich Maeda for useful discussions.

\appendix

\section{Renormalizability of Counter Terms}
\subsection{Extended PCR of Counter Terms}\label{appendix:counter}
Here, we show that, if the extended PCR conditions \eqref{PCR} and \eqref{eq:ePCR} hold, 
the required counter terms also satisfy Eqs.\,\eqref{PCR} and \eqref{eq:ePCR}. 
As in the discussion in Sec.\,\ref{Lfr},
the counter term for an $n$-point diagram with $L$ loops, $P$ propagators and $V$ vertices can be estimated as 
\begin{eqnarray}
\prod_{i=1}^V \prod_{\bar m} \left( \partial^{a_{\bar m}^{i}} \phi \right)
\int (d\omega d^dp)^L \left(\frac{1}{\omega^2-p^{2z}}\right)^P 
\prod_i \left(\prod_m p^{a_{m}^{i}} \right), \label{aploop}
\end{eqnarray}
where $i$ denotes the label of vertices and $m$ and $\bar m$ are the labels for 
$\partial^{a_{m}^i} \phi $ in Eq.\,(\ref{eq:vertex}) corresponding to the internal and external lines at each vertex.
The degree of divergence of the integral can be estimated as 
\begin{eqnarray}
D=z+d-\sum_i \left(z+d-\sum_m \left( a_{m}^i+[\phi]\right)\right). \label{apD}
\end{eqnarray}
Here we have used $[\lambda]= z+d-\sum_{l=1}^n \left( a_l + [\phi] \right)$.
Since we are interested in the divergent contributions, 
we consider the case where $D$ is non-negative. 

The integral in Eq.\,(\ref{aploop}) has sub-leading divergent contributions. 
We expand it with respect to the small value $k/p$ where $k$ is the magnitude of external momenta. 
Expanding the integral, we find that the marginal terms in the counter term have totally $D$ extra spatial derivatives. 
Therefore, the extended PCR conditions for the counter terms are written as 
\begin{eqnarray}
\sum_{i} ~ \sum_{\mbox{\tiny{all $\bar m$}}} ~ (a_{\bar m}^i +[\phi] ) + D &\le& z+d, \label{counterPCR} \\
\sum_{i} \sum_{\mbox{\tiny{part of $\bar m$}}} (a_{\bar m}^i +[\phi] )+ D &<& z+d. 
\label{counterEPCR}
\end{eqnarray}
Substituting Eq.\,(\ref{apD}) into the left hand side of the first inequality, we have 
\begin{eqnarray}
\sum_{i} \sum_{all \  \bar m} (a_{\bar m}^i +[\phi] ) + D =
z + d + \sum_{i} \left[ -(z+d) + \sum_{l=1}^n (a_{l}^i + [\phi] ) \right].
\end{eqnarray} 
This implies that the inequality (\ref{counterPCR}) is met, 
when all the vertices satisfy the conventional PCR condition $\sum_{l=1}^n (a_{l}^i + [\phi] ) \leq z+d$.
Similarly, we can show that the inequality (\ref{counterEPCR}) holds if the extended conditions PCR are satisfied for all vertices.

\subsection{Number of Counter Terms}\label{Apfinite}

If the dimension of a field is non-positive, 
we have an infinite number of terms satisfying the extended PCR. 
For instance, the interaction term $S_{int}=\int dt d^d x\, \phi^n$ satisfies the extended PCR for any $n$. 
It seems that an infinite number of counter terms are required. 
However, we show here that, 
if the number of terms in the bare action is finite, 
we can renormalize the theory with a finite number of counter terms. 

We consider the following generic interaction term
\begin{eqnarray}
S_{m,{\cal D}}=\int dt d^d x \left( \partial_x^{a_1} \phi \right) \dots \left( \partial_x^{a_m} \phi \right)
\left( \partial_x^{b_1} \phi \right)\dots \left( \partial_x^{b_n} \phi \right). \label{geneint}
\end{eqnarray}
Here, we assume that the first $m$ $( \partial_x^{a_j} \phi )$'s have non-positive dimensions 
while the dimensions of the last $n$ $( \partial_x^{b_j} \phi )$'s are positive, 
that is, 
\begin{eqnarray}
[\partial_x^{a_j} \phi] < 0 \hspace{5mm} (j=1,\cdots,m), \hspace{10mm} [\partial_x^{b_j} \phi] > 0 \hspace{5mm} (j=1,\cdots,n).
\end{eqnarray} 
The number ${\cal D}$ denotes the dimension of the product 
$\left( \partial_x^{b_1} \phi \right) \dots \left( \partial_x^{b_n} \phi \right)$,  
that is, for an interaction term $S_{int,m,{\cal D}}$, 
the index $m$ indicates the number of $( \partial_x^{a_j} \phi )$'s with non-positive dimensions 
and ${\cal D}$ shows the total dimension of $( \partial_x^{b_j} \phi )$'s with positive dimensions. 
From the extended PCR conditions, ${\cal D}$ has to satisfy
\begin{eqnarray}
&&{\cal D} \le d+z, \ (\mbox{for}\ m=0) .\label{delta<2}\\
&&{\cal D} < d+z, \ (\mbox{for}\ m\ge 1).  \label{delta<} 
\end{eqnarray} 
The second inequality (\ref{delta<}) can also be rewritten in 
\begin{eqnarray}
{\cal D} \le d+z-1/2, \ (\mbox{for}\ m\ge 1).  \label{delta<3} 
\end{eqnarray} 
Since $b_j$, $d$ and $z$ are all integers, 
$[ \partial_x^{b_j} \phi ]$ has to be a half of integer, that is, $[ \partial_x^{b_j} \phi ] \geq \frac{1}{2}$. 
Because ${\cal D}(\geq n/2)$ is bounded both from above, 
$n$ should also be bounded from above; $n \le 2(d+z)-1$. 
Therefore the form of the part $\left( \partial_x^{b_1} \phi \right) \cdots \left( \partial_x^{b_n} \phi \right)$ is limited.

We consider a generic loop diagram constructed with $p$ interaction terms 
$S_{m_1,{\cal D}_1}$, $\dots$, $S_{m_p,{\cal D}_p}$ with nonzero $m_j$ and 
$q$ interaction terms $S_{m_{p+1},{\cal D}_{p+1}}$, $\dots$, $S_{m_{p+q},{\cal D}_{p+q}}$ with $m_{p+j}=0$. 
The maximum divergence occurs 
when all positive dimensional operators appear as the internal lines 
while negative dimensional operators correspond to external lines. 
Dimension-zero operators can be ether internal lines or external lines. 
Then, the leading order behavior of the UV divergence can be estimated as 
\begin{eqnarray}
&& D=d+z-\sum_{j=1}^p \left( d+z- {\cal D}_j \right) -\sum_{j=1}^q\left( d+z- {\cal D}_{p+j} \right) .
\end{eqnarray} 
In general, other counter terms are required for sub-leading contributions, 
which appear when positive dimensional operators become external lines, 
or when loop diagrams are expanded with respect to external momenta. 
However, the loop integrals become finite
if enough many positive dimensional operators become external lines,
or if there are enough higher order contributions from the expansion. 

We consider a divergent loop diagram in which some positive dimensional operators appear in the external lines. 
Suppose that $S_{m_c,{\cal D}_c}$ is the corresponding counter term 
and ${\cal D}_c$ is the sum of dimensions of the positive dimensional operators. 
The degree of divergence is read as 
\begin{eqnarray}
D = d+z - {\cal D}_c - \sum_{j=1}^p \left( d+z- {\cal D}_j \right) -\sum_{j=1}^q\left( d+z- {\cal D}_{p+j} \right) .
\end{eqnarray}
Since $D \geq 0$ for divergent diagrams, counter terms cannot have arbitrarily large ${\cal D}_c$. 
The bound for ${\cal D}_c$ can be found from inequalities (\ref{delta<2}) and (\ref{delta<3}) as
\begin{eqnarray}
0 \leq D \le \left\{
  \begin{array}{lc}
     {\cal D}_k-{\cal D}_c -\frac{p-1}{2} & (\mbox{for} \ k \le p\ \mbox{and} \ p \ge 1)  \\
    {\cal D}_k-{\cal D}_c -\frac{p}{2} & (\mbox{for} \ k \ge p+1\ ) \\
  \end{array}
\right. , \label{Dc}
\end{eqnarray}
where we have replaced ${\cal D}_j$ with their maximum values except for ${\cal D}_{j=k}$. 
Thus, ${\cal D}_c$ has to be smaller than or equal to ${\cal D}_k$, 
and hence all counter terms satisfy the extended PCR condition, 
as we have seen in Appendix \ref{appendix:counter}. 
  
The value of $p$ is also suppressed by inequality (\ref{Dc}). 
Thus, $m_c$ is bounded from above as
\begin{eqnarray}
m_c \leq \sum_{i=1}^{p_{max}} m_i \leq m_{max} p_{max},
\end{eqnarray}
where $m_{max}$ and $p_{max}$ are the maximum value of $m_i$ and $p$, respectively. 
When a counter term $S_{m_c, {\cal D}_c}$ is not included in the original bare action, 
it has to be added to the modified action.
Then, if the counter term has $m_c > m_{max}$, 
the maximum value $m_{max}$ is also modified as $m_{max}^{modified} = m_c$. 

In the modified theory, the added counter terms may give rise to new divergences and 
we may need to iteratively add new counter terms with larger $m_c$. 
We can show that this operation terminates after finite iterations as follows.
Counter terms with $m_c > m_{max}$ are required to cancel divergences from diagrams with $p \geq 2$. 
We can see from inequality \eqref{Dc} that ${\cal D}_c \geq {\cal D}_k - \frac{1}{2}$.  
This implies that ${\cal D}_c$ decreases by the iterative operation, 
but ${\cal D}_c$ has to be positive.
Thus, the operation stops and 
the counter terms in the final operation have only non-positive dimensional operators. 
As a result, $m_c$ is bounded from above.

We check that loop contributions constructed 
with the operator having only non-positive dimensional operators require a finite number of counter terms.
We consider operators with $q=0$, i.e. all operators have non-positive dimensions. 
By assumption, the dimension of each coupling constant must be larger than $d+z$, that is, 
\begin{eqnarray}
[\lambda_j] > d+z.  \label{lamj} 
\end{eqnarray}
The degree of divergence of a loop diagram can be estimated as 
\begin{eqnarray}
D=[\lambda_c]-\sum_{j=1}^n [\lambda_j] ,
\end{eqnarray}
where $[\lambda_c]$ is the dimension of the coupling constant in the required counter terms.
If the counter terms are required, 
$D$ is equal to or larger than zero, 
and then from (\ref{lamj}), $[\lambda_c]$ is suppressed as 
\begin{eqnarray}
[\lambda_c] \le [\lambda_j] .
\end{eqnarray}
Therefore, the number of counter terms for loop diagrams constructed with $q=0$ is finite.

Since ${\cal D}_c$ and $m_j$ are bounded from above 
and a finite number of counter terms are required for the loop diagram constructed with $q=0$,
the number of counter terms are finite if the original action has a finite number of terms.

\section{Evaluation of Phase Space Integrals}\label{appendix:integral}
In this section, we derive the asymptotic forms of the phase space integrals used in this paper. 
We assume that the dispersion relation takes the form
\begin{eqnarray}
E ~=~ f(p^2) ~=~ p^{2z} + \eta^{2z} \sum_{i=0}^{n-1} c_{2i} \left( \frac{p}{\eta} \right)^{2i},
\end{eqnarray}
where $\eta$ is a scale parameter with $[\eta]=1$ and $c_{2i}$ are dimensionless parameters.
Let us consider the integral of the form
\begin{eqnarray}
X ~=~ \int_{I} \frac{d^d p_1}{2 E_1} \frac{d^d p_2}{2E_2} \delta(E_1 + E_2 - E) \delta^d( \mathbf p_1 + \mathbf p_2 - \mathbf P) 
\, F(\mathbf p_1, \mathbf p_2,  \mathbf P, \eta),
\end{eqnarray}
where $F(\mathbf p_1, \mathbf p_2, \mathbf P, \eta)$ is a function of the momenta and the scale parameter
which does not depend on other dimensionful parameters. 
We focus on the subspace of two particle Hilbert space where the total energy and momentum are related as
\begin{eqnarray}
E = P^{z} + \mathcal O(P^{z-1}).
\end{eqnarray}

Let us first consider the integral corresponding to the state $|\alpha \rangle$,
for which the domain of integration is given by
\begin{eqnarray}
I~=~I_\alpha~=~\big\{ \, (\mathbf p_1, \mathbf p_2) \in \mathbb R^d \times \mathbb R^d ~\big|~ \big||\mathbf p_1|-|\mathbf p_2|\big| \leq P/2 \, \big\}.
\end{eqnarray}
Changing the integration variable as $\mathbf p_i = P \mathbf u_i$ and assuming that $[F]=a$, i.e.
\begin{eqnarray}
F(\mathbf p_1, \mathbf p_2, \mathbf P, \eta) = P^a \, F(\mathbf u_1, \mathbf u_2, \hat{\mathbf P}, \eta/P), \hspace{10mm}
(\hat{\mathbf P} = \mathbf P / P),
\end{eqnarray}
we find that for small $\delta \equiv \eta/P$, the asymptotic form of $X$ is given by
\begin{eqnarray}
X = P^{d-3z+a} \bigg[ \int_{I'_\alpha} \frac{d^d u_1 d^d u_2}{4 u_1^z u_2^z} \delta(u_1^z + u_2^z - 1) \delta^d( \mathbf u_1 + \mathbf u_2 - \hat{\mathbf P}) F(\mathbf u_1, \mathbf u_2, \hat{\mathbf P}, 0) + \mathcal O(\delta) \bigg], 
\label{eq:X}
\end{eqnarray}
where the domain of integration is 
$I_\alpha' = \big\{ \, (\mathbf u_1, \mathbf u_2) \in \mathbb R^d \times \mathbb R^d ~\big|~ \big||\mathbf u_1|-|\mathbf u_2| \big| \leq 1/2 \, \big\}$.
Since the leading term in Eq.\,\eqref{eq:X} is independent of $P$, 
the asymptotic form of $X$ can be estimated as
\begin{eqnarray}
X ~\approx~ const. \times P^{d-3z+a}.
\end{eqnarray}

Next, let us consider the integral corresponding to the state $|\beta \rangle$,
for which the domain of integration is given by
\begin{eqnarray}
I~=~I_\beta ~=~\big\{ \, (\mathbf p_1, \mathbf p_2) \in \mathbb R^d \times \mathbb R^d ~\big|~ |\mathbf p_1| \leq \epsilon \equiv  c \eta \, \big\},
\end{eqnarray}
where $c$ is a dimensionless parameter.
Integrating over $\mathbf p_2$ and changing the integration variable as $\mathbf p_1 = \eta \boldsymbol v$,
we find that the asymptotic form of the energy conservation factor is given by
\begin{eqnarray}
E_1 + E_2 - E &\approx& \eta P^{z-1} \left[ - z \, \hat{\mathbf P} \cdot \boldsymbol v + \mathcal O(\delta) \right], \hspace{10mm} (\mbox{for $z >1$}).
\end{eqnarray}
Assuming that the asymptotic form of the function $F$ takes the form
\begin{eqnarray}
\eta^{-a} F(\mathbf p_1, \mathbf p_2, \mathbf P, \eta) \approx \frac{P^{b}}{\eta^b} \left( v^q \tilde F(\theta) + \mathcal O(\delta) \right), \hspace{5mm} \left( [F]=a, ~~ \hat{\mathbf P} \cdot \boldsymbol v = v \cos \theta \right),
\end{eqnarray}
we can approximate the integral as
\begin{eqnarray}
X &=& \eta^a \int_{|\boldsymbol v| < c} d^d v 
\frac{P^b / \eta^{b} \left( v^q \tilde F(\theta) + \mathcal O(\delta) \right)}{\sqrt{v^{2z} + \sum c_{2i} v^{2i}} \sqrt{P^{2z} + \mathcal O(\delta)}} 
\delta\left( \eta P^{z-1} \left[ - z \, \hat{\mathbf P} \cdot \boldsymbol v + \mathcal O(\delta) \right] \right).
\end{eqnarray}
Therefore, the leading term in $X$ is given by
\begin{eqnarray}
X &\approx& const. \times P^{1-2z+b}, \hspace{10mm} 
\left( const. = \eta^{a-b-1} \int_{|\boldsymbol v| < c} \frac{v^q \tilde F(\theta) d^d v}{\sqrt{v^{2z} + \sum c_{2i} v^{2i}}} \right).
\end{eqnarray}

\end{document}